\documentclass[journal=jctcce,manuscript=article,layout=twocolumn]{achemso}

\usepackage[version=3]{mhchem} 
\usepackage{xcolor}
\usepackage{booktabs}  
\usepackage{caption}   
\usepackage{multirow}
\usepackage{subfigure}
\usepackage{array}
\usepackage{threeparttable}
\expandafter\def\csname ver@fixltx2e.sty\endcsname{2016/01/03 v2.0a}
\usepackage{dblfloatfix}

\author{Zehao Zhou}
\email{zhouzehao@bza.edu.cn}
\affiliation[agc]
{Zhongguancun Academy, Beijing, China}

\author{Xiaojie Wu}
\affiliation[bytedance]
{Bytedance Seed, San Jose}

\author{Yanheng Li}
\affiliation[PKU]
{The College of Chemistry and Molecular Engineering, Peking University, Beijing, China}
\alsoaffiliation[bytedance]
{Bytedance Seed, Beijing}

\author{Xinran Wei}
\affiliation[agc]
{Zhongguancun Academy, Beijing, China}

\author{Cheng Fan}
\affiliation[PKU]
{The College of Chemistry and Molecular Engineering, Peking University, Beijing, China}
\alsoaffiliation[bytedance]
{Bytedance Seed, Beijing}

\author{Fusong Ju}
\affiliation[agc]
{Zhongguancun Academy, Beijing, China}

\author{Qiming Sun}
\email{qiming.sun@bytedance.com}
\affiliation[bytedance]
{Bytedance Seed, Seattle}

\author{Yi Qin Gao}
\email{gaoyq@pku.edu.cn}
\affiliation[NPKU]
{New Cornerstone Science Laboratory, The College of Chemistry and Molecular Engineering, Peking University, Beijing, China}
\alsoaffiliation[WLU]
{West Lake University, Hangzhou, China}

\title[GPU-TDDFT-risp for Large Molecules]
  {GPU Accelerated Minimal Auxiliary Basis Approach TDDFT
  for Large Organic Molecules}

\begin{document}

\begin{tocentry}





\centering
\includegraphics[width=\linewidth]{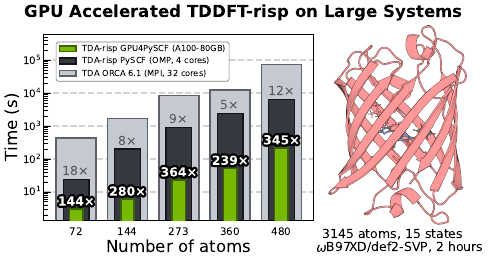}

\end{tocentry}

\begin{abstract}
We introduce a GPU-accelerated implementation of time-dependent density
functional theory with the minimal auxiliary basis approach (TDDFT-risp) in
GPU4PySCF, together with large system demonstrations carried out using the
Tamm--Dancoff approximation (TDA-risp). The method combines GPU-accelerated
three-center integral evaluation, tensor contractions, exchange-space
truncation, omission of hydrogen atoms from the auxiliary basis, and a host
memory assisted Davidson solver. On the EXTEST42 benchmark set, a conservative
40 eV exchange cutoff yields excitation-energy errors relative to standard TDA
of about 0.03--0.05 eV for low-lying states. For systems of 300 to 3000 atoms,
we demonstrate that TDA-risp calculations of 15 low-lying excited states with
$\omega$B97XD/def2-SVP complete on a single A100 GPU with wall times ranging from
minutes to hours. These results position GPU-TDDFT-risp as a practical route
toward excited-state calculations for large organic and biomolecular systems
with thousands of atoms.
\end{abstract}

\section{Introduction}
Excited-state calculations on molecular systems with thousands of atoms are
demanded by a growing range of applications in chemistry and materials science.
For example, fluorescent proteins (typically 2000--5000 atoms) underpin
bioimaging and biosensing. Their excited-state proton transfer, conical
intersections, and internal conversion govern emission wavelength and quantum
yield, and rational design of red-shifted or brighter probes relies on reliable
excited-state geometries and gradients~\cite{list2019jctc,bourneworster2024jcp}.
In photosystem II and related reaction centers, the protein matrix imposes
asymmetric electrostatic fields that are the primary cause of excitation
asymmetry, and omitting the full environment leads to qualitatively wrong
assignments~\cite{sirohiwal2020jacs}. Organic photovoltaics, OLEDs, and covalent
organic frameworks involve aggregates or crystallites with thousands of atoms,
where exciton migration, charge-transfer states, and triplet--singlet pathways
determine efficiency and device metrics~\cite{craciunescu2023jctc}. In all these
cases, the relevant electronic structure extends over thousands of atoms,
motivating methods that can treat entire systems at the first-principles level.

Simulating excited-state properties using linear-response time-dependent density
functional theory (TDDFT)\cite{casida1995} or other wavefunction-based
\textit{ab initio} methods becomes particularly demanding for systems with
thousands of atoms, due to the prohibitive computational expense associated with
the construction and diagonalization of large response matrices. Significant
progress has recently been made through massive parallelization on modern GPU
architectures. For example, Kim and co-workers implemented a highly efficient
GPU-accelerated TDDFT code capable of running on 256 NVIDIA A100 GPUs. They
successfully computed the \ce{S_1} excitation energy and gradient of the
4200-atom green fluorescent protein chromophore environment
\cite{kim2024jctc}, demonstrating the feasibility of pushing conventional TDDFT
to biologically relevant system sizes.

In parallel, several reduced-cost and semi-empirical strategies have been
developed to extend excited-state calculations to large systems. The
sTDA\cite{grimme2013simplified} and xTB-sTDA\cite{grimme2016jcp} methods achieve
errors on the order of 0.3--0.5~eV for excitation energies with dramatically
reduced cost, and the sTDDFT family has been extended to electronic circular
dichroism (ECD) spectra of large biomolecules\cite{bannwarth2014CaTCa}. The
TDDFT+TB approach combines DFT-based local electronic structure with
tight-binding scaling\cite{ruger2016jcp}. On the tight-binding side, DIALECT
implements TD-DFTB in Rust and has been applied to assemblies of up to 2070
atoms\cite{einsele2025jpca}, while PYSEQM~2.0 employs PyTorch-based GPU
acceleration for NDDO-type excited states on nearly 1000
atoms\cite{athavale2025jctc}. The recently proposed atomic density-based
tight-binding (aTB) model is a semiempirical method with explicit second- and
third-order expansions of the exchange-correlation term and has also been
applied to excited-state calculations\cite{zhang2025jctcb}. These semiempirical
methods are computationally very efficient. Most of them, however, have error of
0.2--0.5 eV relative to \textit{ab initio} TDDFT with hybrid and range-separated
hybrid (RSH) functionals. In this work, our goal is to retain the negligible
errors compared to \textit{ab initio} TDDFT with hybrid and range-separated
hybrid (RSH) functionals on large systems while keeping the cost manageable on a
single GPU card.

The semi-empirical TDDFT-ris method provides a promising starting point with
\textit{ab initio} accuracy\cite{zhou2023jpcl,pu2026jctc}. By replacing the
standard auxiliary basis with a compact $s/p$-type orbitals, TDDFT-ris reduces the formal
prefactor of both the Coulomb (J) and exchange (K) terms. Fragment-based
approach combined with TDDFT-ris has also been proposed to reduce the
computational scaling\cite{lv2026c}. Fragmentation methods are highly effective
for systems dominated by local excitations, whereas a full-system treatment
remains important for charge-transfer states. However, the first-generation
implementation remains far from sufficient for large systems in practice. Its
scalability is typically limited to around 100 atoms because it stores
$\mathcal{O}(N^3)$ molecular orbital (MO)-basis three-center electron repulsion
integral (ERI) tensors for both J and K, which quickly become a memory
bottleneck as system size grows. In addition, once these tensors are built, each
Davidson matrix--vector product is dominated by tensor contractions, and the CPU
execution is constrained by memory bandwidth and limited thread-level
parallelism.

In this work, we remove these bottlenecks and make TDDFT-ris practical for much
larger systems through algorithmic improvements together with a GPU-oriented
implementation in GPU4PySCF. Building on the same minimal auxiliary basis
approximation strategy, wwo key improvements enable this scalability: (i) an
efficient on-the-fly J engine that avoids storing MO-basis J tensors by
evaluating the J contribution directly in the atomic orbital (AO) basis, and
(ii) an energy-window truncation strategy that reduces the effective prefactor
of K tensors. As a result, stored ERIs are no longer the dominant memory cost.
Davidson expansion vectors become the main memory term. To handle systems beyond
device capacity, K tensors and vectors are kept in host memory and streamed to
the GPU in chunks for the contraction kernels. Our current implementation
targets a single GPU, but distributing these chunks across multiple GPUs is a
straightforward extension that we plan to explore in future work.

The paper is organized as follows: Section~\ref{sec:method} reviews the
TDDFT-risp formalism and details. Section~\ref{sec:implementation} describes the
implementation of the on-the-fly Coulomb evaluation, exchange-space truncation,
and the hydrogen-free auxiliary basis. Section~\ref{sec:results} validates the
accuracy on the EXTEST42 benchmark set\cite{zhou2023jpcl} with hybrid and RSH
functionals, profiles wall times and memory usage for systems up to ${\sim}3000$
atoms, and presents illustrative excited-state analyses of fluorescent proteins
and a photosystem~II model.

\section{TDDFT-risp method}\label{sec:method}

Within linear response TDDFT, excitation
energies are obtained by solving the symplectic eigenvalue equation
\begin{equation}
\label{eq:tddft}
  \left(
  \begin{matrix}
  \mathbf{A} & \mathbf{B} \\
  \mathbf{B} & \mathbf{A}
  \end{matrix}\right)
  \left(\begin{matrix}
  \mathbf{X}\\
  \mathbf{Y}
  \end{matrix}\right)
  =
  \left(\begin{matrix}
  \mathbf{1} & \mathbf{0} \\
  \mathbf{0} & \mathbf{-1}
  \end{matrix}
  \right)
  \left(\begin{matrix}
  \mathbf{X}\\
  \mathbf{Y}
  \end{matrix}\right)
  \mathbf{\Omega},
\end{equation}
where
\begin{subequations} \label{eq:hessian}
\begin{align}
    (A + B)_{ia,jb} =& (\varepsilon_{a} - \varepsilon_i) \delta_{ab}\delta_{ij} + 2(ia|jb) + 2f^\text{xc}_{iajb} \nonumber\\
       & - c_x[(ib|ja)+(ij|ab)],\\
    (A - B)_{ia,jb} =& (\varepsilon_{a} - \varepsilon_i) \delta_{ab}\delta_{ij}  \nonumber\\
    & + c_x[(ib|ja)-(ij|ab)],
\end{align}
\end{subequations}
$\varepsilon_p$ is the KS eigenvalue associated with KS orbital $\phi_p$,
$f^\text{xc}_{iajb}$ is a matrix element of the semilocal exchange-correlation
(XC) kernel, $(pq|rs)$ is a four-center ERI, and $c_x$ is the fraction of Hartree-Fock
exchange. In Eq.~\eqref{eq:hessian}, labels $i,j$ refer to occupied MOs
while $a,b$ refer to virtual MOs.

In TDDFT-risp, the XC kernel in Eq.~\eqref{eq:hessian} is neglected and the
expensive ERIs are approximated with a minimal auxiliary basis expansion, as
introduced in the previous work.~\cite{zhou2023jpcl}. The dominant cost is the matrix--vector
product in Davidson iterations. The J term is evaluated on-the-fly in the AO
basis, exploiting sparsity to achieve $\mathcal{O}(N^2)$--$\mathcal{O}(N^3)$
effective scaling. The K term is formally $\mathcal{O}(N^4)$, but the MO-space
energy-window truncation dramatically reduces the prefactor by replacing
$N_\text{occ}$ and $N_\text{virt}$ with $N_\text{occK'}$ and
$N_\text{virtK'}$~\cite{zhou2023jpcl}. The ERIs are approximated using the
resolution-of-the-identity (RI) technique,
\begin{equation}
\label{eq:TDDFT-axs_approx_integral}
    (pq|rs) \approx \sum_{PQ}(pq|P)(P|Q)^{-1}(Q|rs),
\end{equation}
where $(P|Q)$ and $(pq|P)$ are two-center and three-center two-electron
repulsion integrals, respectively. $p,q,r,s$ label generic
MOs \cite{krause2017JCC}. $P,Q$ label atom-centered auxiliary basis, which are a
set of minimal primitive Gaussian basis functions without contraction,
\begin{equation}\label{eq:gr}
    g(r) = x^{l_x} y^{l_y} z^{l_z} \exp(-\frac{\theta}{R_A^2}r^2),
 \end{equation}
where $\theta$ is a global parameter tuned to be 0.2 in the previous
work~\cite{zhou2023jpcl}. In this work, we use the same value of $\theta$ for
all systems. $R_A$ is the semi-empirical radius of atom $A$\cite{ghosh2008JMS}.
The sum of $l_x, l_y, l_z$ is the angular momentum $l$ of the primitive Gaussian
basis function, e.g., $l=0$ for $s$-type, $l=1$ for $p$-type, $l=2$ for
$d$-type, etc. Eq.~\eqref{eq:TDDFT-axs_approx_integral} is also employed in
TDDFT-as, but only for the Coulomb contribution, whereas in TDDFT-risp this is
extended to hybrid and RSH functionals.

The $n^\text{th}$-excited state oscillator strength and rotatory strength are given by
\begin{align}
     f_{0n} &= \frac{2}{3} \Omega_n \boldsymbol{\mu}_{0n}  \boldsymbol{\mu}_{n0},  \label{eq:oscillator_strength} \\
     R_{0n} &= \frac{1}{2} \boldsymbol{\mu}_{0n} \boldsymbol{m}_{0n}, \label{eq:rotator_strength}
\end{align}
where the electric dipole transition moment $\boldsymbol{\mu}_{0n}$ and the magnetic dipole transition
moment $\boldsymbol{m}_{0n}$ in dipole length formalism are given by
\begin{align}
    \boldsymbol{\mu}_{0n} &= \sum_{ia}  \boldsymbol{r}_{ia} (X+Y)_{nia},  \label{eq:transition} \\
    \boldsymbol{m}_{0n}  &= \sum_{ia} (\boldsymbol{r}\times  \boldsymbol{\nabla})_{ia} (X-Y)_{nia}. \label{eq:transition_magnetic}
\end{align}

Equations~\eqref{eq:transition}--\eqref{eq:transition_magnetic} use the full
linear-response amplitudes $(\mathbf{X},\mathbf{Y})$\cite{bannwarth2014CaTCa}. In the Tamm--Dancoff
approximation (TDA)\cite{hirata1999cpl}, one sets $\mathbf{Y}=\mathbf{0}$ and
replaces the symplectic eigenvalue problem of Eqs.~\eqref{eq:tddft} and
\eqref{eq:hessian} by a single Hermitian problem in $\mathbf{X}$ alone. The
TDA-risp eigenvalue equation reads
\begin{equation} \label{eq:TDA}
    \mathbf{A} \mathbf{X} = \mathbf{X}\mathbf{\Omega},
\end{equation}
with
\begin{equation} \label{eq:TDA-risp}
    A_{ia,jb} = (\varepsilon_{a} - \varepsilon_i) \delta_{ab}\delta_{ij} + 2(ia|jb) - c_x(ij|ab).
\end{equation}
For pure functionals, $c_x = 0$.
Both TDDFT-risp and TDA-risp are available in GPU4PySCF and
PySCF\cite{sun2018wcms,sun2020jcp}.

\section{Implementation details}\label{sec:implementation}
This section describes four key aspects of the implementation that together make
thousand-atom calculations feasible on a single GPU:
(i) the J term is evaluated on-the-fly in the AO basis, eliminating
MO-basis J tensor storage (Section~\ref{sec:J});
(ii) an energy-window truncation and hydrogen-free auxiliary basis reduces the prefactor of the K term (Section~\ref{sec:K});
(iii) a memory-aware Davidson solver streams data between host and device
memory and triggers restarts before host memory is exhausted (Section~\ref{sec:davidson}).

\subsection{On-the-Fly Coulomb Evaluation}\label{sec:J}
The Coulomb contribution is evaluated in the AO representation rather than building MO-basis ERIs:
\begin{align}
\label{eq:J_mia}
J_{mia} = \sum_{\mu \nu \kappa \lambda} C_{\mu i} C_{\nu a} (\mu \nu|\kappa \lambda)
D^m_{\kappa \lambda},
\end{align}
where the AO-basis transition density is
\begin{align}
D^m_{\kappa \lambda} = \sum_{jb} C_{\kappa j} C_{\lambda b} X_{mjb}.
\end{align}
Using RI in the Coulomb metric,
\begin{align}
(\mu \nu|\kappa \lambda) \approx \sum_{PQ} (\mu \nu|P) (P|Q)^{-1}
(Q|\kappa \lambda),
\end{align}
the contraction is carried out in the following order:
\begin{align}
R_{mQ} &= \sum_{\kappa \lambda} (Q|\kappa \lambda) D^m_{\kappa \lambda}
\equiv \sum_{z} I_{Qz} D^m_{z}, \\
\tilde{R}_{mP}
&\equiv \sum_{Q} (P|Q)^{-1} R_{mQ}, \\
U^m_{\mu \nu}
&\equiv \sum_{P} (\mu \nu|P) \tilde{R}_{mP}, \\
J_{mia}
&= \sum_{\mu \nu} C_{\mu i} C_{\nu a} U^m_{\mu \nu}.
\end{align}
Here, $z$ indexes the screened non-zero AO pairs $(\kappa,\lambda)$, and
$I_{Qz}$ stores the corresponding lower-triangular elements of
$(Q|\kappa \lambda)$. In practice, $I_{Qz}$ is generated on-the-fly and the
auxiliary index $Q$ is processed in chunks according to available GPU memory.
For each state $m$, the final AO-to-MO back-transformation requires only two
tensor contractions. Since $N_\text{states} \ll N_\text{aux}$ in typical
calculations (e.g., 20 states versus $N_\text{aux}$), iterating over
$N_\text{states}$ instead of $N_\text{aux}$ is computationally efficient.

\subsection{Exchange-Space Truncation}\label{sec:K}
In the context of standard TDDFT with hybrid and RSH functionals, the exchange
term $K_{mia} = \sum_{jb} (ij|ab) X_{m jb}$ is $\mathcal{O}(N^4)$ scaling and
thus the computational bottleneck.

Unlike the Coulomb term, the exchange term cannot exploit the transition density
in AO-basis sparsity. Instead, intermediate tensors $\mathbf{T}_{ij}^P$ and
$\mathbf{T}_{ab}^P$ are built and stored in memory. Here we use two strategies
to reduce the size of the exchange tensors: (i) truncation of low-lying
occupied and high-lying virtual MOs by an energy window, and (ii) omission of
auxiliary basis functions on hydrogen atoms. These two approximations define the
main exchange-side acceleration strategy that is validated in the Results and
discussion section before being combined with the large-system Davidson
algorithm.

In the previous preconditioning studies\cite{zhou2024jctc}, a simple yet
effective MO truncation strategy was employed to control computational costs.
This ensured that the preconditioning step consumed no more than 1\% of one ab
initio TDDFT matrix-vector product cost, while not compromising the
preconditioning effectiveness. Specifically, virtual MOs with orbital energies
$\epsilon \geq 40$ eV were truncated. Similarly, occupied MOs with orbital
energies $\epsilon \leq -40$ eV were also truncated. These measures
yielding reduced dimensions $N_{\text{occK}'} < N_{\text{occ}}$ and
$N_{\text{virtK}'} < N_{\text{vir}}$.
Here, we utilize this MO truncation strategy and the primary goal is to preserve the accuracy of low-lying
excitation energies and oscillator strengths, rather than the preconditioning
efficiency. The quantitative accuracy tradeoff associated with the truncation
threshold is examined later on the EXTEST42 benchmark set.

A further approximation is to omit auxiliary basis functions on hydrogen atoms.
The first excited state of atomic hydrogen lies at 10.2~eV, far above the
typical excitation window of organic chromophores ($\leq 5.5$~eV), so hydrogen
orbitals seldom contribute significantly to low-lying excited states. In
hydrogen-rich organic systems, excluding hydrogen from the auxiliary basis
functions on hydrogen in the exchange term can in principle reduce the K-term
cost by roughly a factor of two. The quantitative impact on excitation-energy
accuracy, including a charge-transfer example, is evaluated in the Results and
discussion section. In the production setting used below, hydrogen is excluded
from both the Coulomb and exchange auxiliary bases.

\begin{figure}[!htbp]
    \centering
    \includegraphics[width=0.9\linewidth]{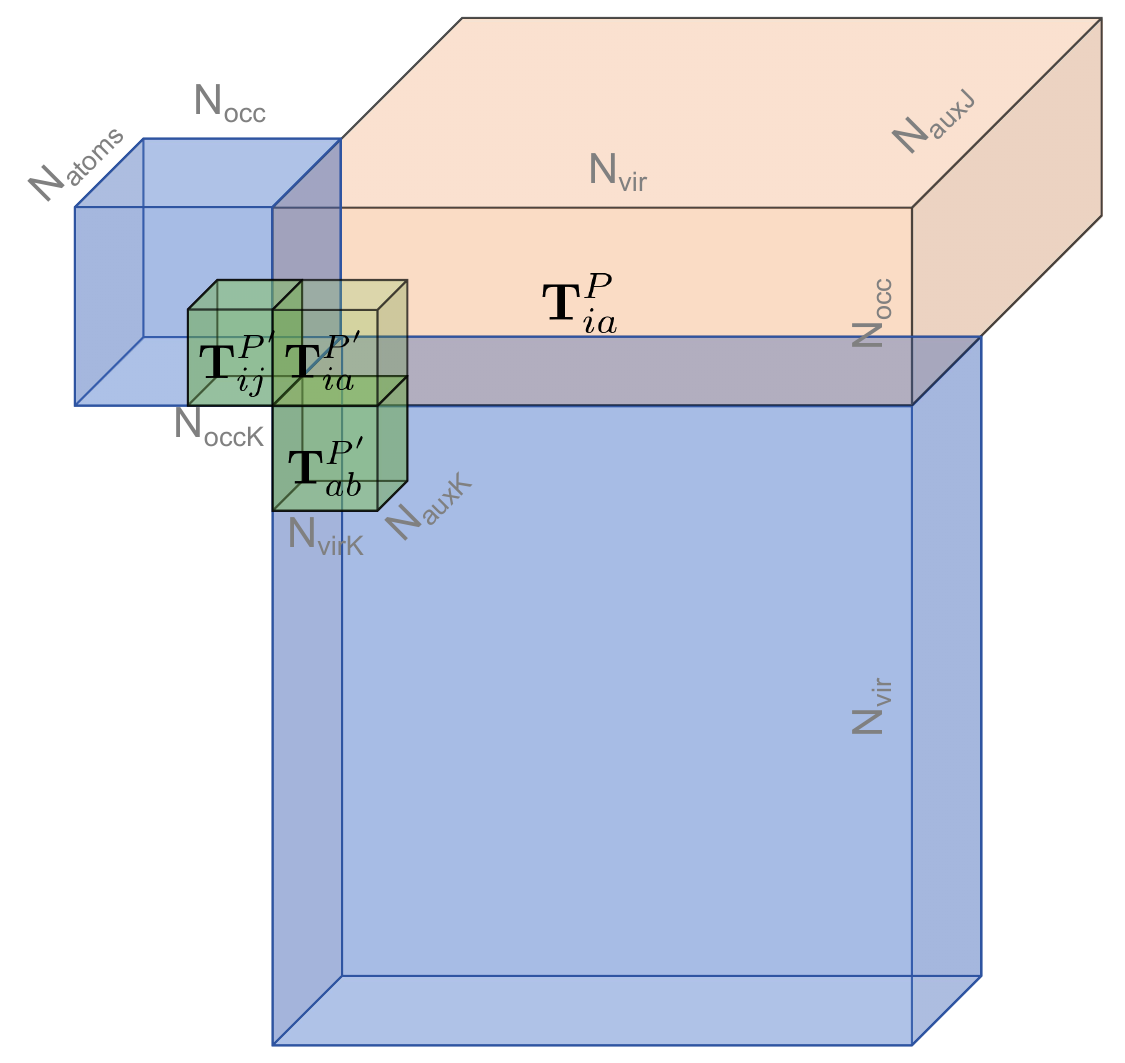}
    \caption{\label{fig:tensors_shape}
    MO-basis three-center ERI tensors. Orange $\mathbf{T}_{ia}^P$ for Coulomb, green
    $\mathbf{T}_{ij}^{P'}$ and $\mathbf{T}_{ab}^{P'}$ for exchange, and $\mathbf{T}_{ia}^{P'}$ exclusively for full TDDFT
    exchange. $N_\text{auxK}$ and $N_\text{auxJ}$ are the number of auxiliary
    basis functions for the exchange and Coulomb terms, respectively.
    $N_\text{occK'}$ and $N_\text{virtK'}$ are the number of occupied and
    virtual MOs after truncation, thus less than the standard amount
    $N_\text{occ}$ and $N_\text{vir}$.
    }
\end{figure}

Beside above truncation strategies, the exchange tensor is built with only one
$s$-type Gaussian fitting function per non-hydrogen atom. In contrast, the
Coulomb term includes an additional $p$-type Gaussian on non-hydrogen centers
for higher spectral accuracy. Namely, $N_\text{auxJ} = 4 N_\text{auxK}$.
Figure~\ref{fig:tensors_shape} illustrates the structure of the key three-center
ERI tensors involved. The orange tensor $\mathbf{T}_{ia}^P$ corresponds to the
Coulomb tensor. The green tensors $\mathbf{T}_{ij}^{P'}$ and
$\mathbf{T}_{ab}^{P'}$ represent the exchange tensors, which appear in hybrid and
RSH functionals. Notably, the tensor $\mathbf{T}_{ia}^{P'}$ is required only in
the full TDDFT formalism and is omitted in the TDA.

Here and in the equations that follow, the indices $i,j$ run over the
truncated occupied space ($N_\text{occK'}$) and $a,b$ over the truncated
virtual space ($N_\text{virtK'}$) introduced above. The auxiliary index
$P$ runs over $N_\text{auxK}$.
The exchange term in the matrix-vector product is given by
\begin{align}
    \label{eq:K_mia}
    K_{mia}
    &= \sum_{jb} (ij|ab) X_{mjb}.
    \end{align}
Using RI for the exchange ERI, we evaluate
\begin{align}
K_{mia}
&\approx \sum_{jb} \sum_{P} T^P_{ij}T^P_{ab}X_{mjb},
\end{align}
where two rank-3 MO basis ERI tensors are precalculated and stored in memory,
\begin{align}
T^P_{ij} &\equiv \sum_{Q}(ij|Q)(Q|P)^{-1},\\
T^P_{ab} &\equiv \sum_{Q}(ab|Q),
\end{align}
which leads to the $\mathcal{O}(N^3)$ memory scaling for the exchange term.
By iterating over chunks in $N_\text{auxK}$, a rank-4 intermediate tensor $M^P_{mja}$ is unavoidable,
\begin{align}
    M^P_{mja} &\equiv \sum_b T^P_{ab}X_{mjb},\\
    K_{mia} &= \sum_{P} \sum_{j}T^P_{ij}M^P_{mja},
\end{align}
which leads to a formal $\mathcal{O}(N^4)$ computation cost for the exchange
term. In practice, the MO-space truncation described above replaces
$N_\text{occ}$ and $N_\text{virt}$ by the much smaller $N_\text{occK'}$ and
$N_\text{virtK'}$, dramatically reducing the prefactor while leaving the
formal scaling unchanged. As a concrete example, for the 3000-atom 6a25 system
with a 16~eV threshold and def2-SVP basis, the K-term tensors
$\mathbf{T}_{ij}^P$ and $\mathbf{T}_{ab}^P$ together require only
${\sim}45$~GB (Table~\ref{tab:large_systems_profiling}), which is considerably
smaller than the Davidson expansion vectors
$\mathbf{V}$ and $\mathbf{W}$ (${\sim}149$~GB for 15 states). Thus, for
systems up to ${\sim}3000$ atoms, the exchange tensor storage is no longer the
memory bottleneck. The Davidson expansion vectors are.

In the RSH functional, the exchange-type ERI splits into two short-range and
long-range parts
\begin{align}
\label{eq:RSH}
(\mu \nu| \kappa \lambda) = \underbrace{(\mu \nu|\frac{1}{r_{12}}-\hat{J}|\kappa \lambda)}_\text{short range}
    + \underbrace{(\mu \nu|\hat{J}|\kappa \lambda)}_\text{long range},
\end{align}
where the $\omega$-attenuated range-separated Coulomb operator $\hat{J}$ is defined as
\begin{align}
\hat{J} = \frac{\alpha + \beta \cdot \text{erf}(\omega r_{12})}{r_{12}}.
\end{align}
The first term in Eq.~\eqref{eq:RSH} is the short-range contribution that is
treated at the semilocal DFT level in the parent response kernel and is not
explicitly represented in the present TDDFT-risp exchange build. The second
term corresponds to the long-range Hartree--Fock exchange
and is approximated using the same RI technique
\begin{align}
\label{eq:K_mia_RSH}
    (\mu \nu| \kappa \lambda)^\text{LR} &=  \sum_{PQ} (\mu \nu|\hat{J}|P)(P|\hat{J}|Q)^{-1}(Q|\hat{J}|\kappa \lambda)
\end{align}
The EXTEST42 benchmarks reported below include RSH functionals
($\omega$B97XD, $\omega$B97X-D3BJ, $\omega$B97MV), so the overall accuracy of
the long-range exchange approximation is assessed together with the other
approximations rather than in isolation.

\subsection{Large-System Davidson Diagonalization}\label{sec:davidson}

\begin{figure}[!htbp]
    \centering
    \includegraphics[width=\linewidth]{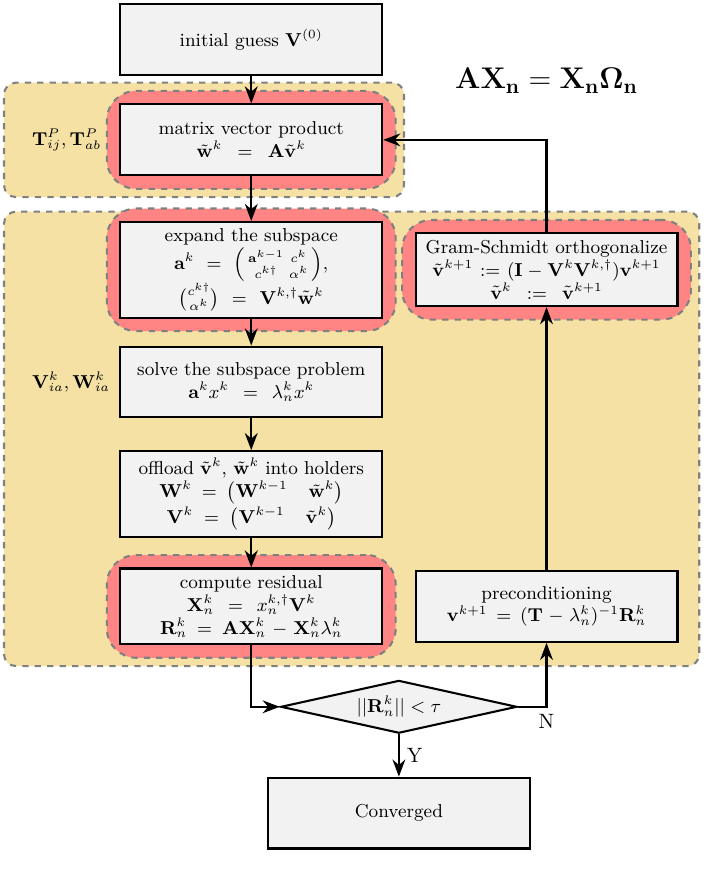}
    \caption{\label{fig:flow_chart_raw}
    The Davidson diagonalization flow chart with GPU acceleration. The yellow
    areas indicate major CPU memory-resident quantities. The red areas highlight
    intensive GPU computations and CPU--GPU data transfers.}
\end{figure}

Figure~\ref{fig:flow_chart_raw} illustrates the Davidson
diagonalization workflow. The yellow areas highlight the dominant memory
consumers: the three-center ERIs, $\mathbf{T}_{ij}^P$ and
$\mathbf{T}_{ab}^P$, required by the exchange term in the matrix--vector
product every iteration. The trial and projection vectors, $\mathbf{V}_{ia}^k$ and
$\mathbf{W}_{ia}^k$, accumulated during the Davidson subspace expansion. The
red areas denote compute-intensive steps involving GPU matrix multiplications
and data transfers between CPU and GPU memory.

The implementation is intentionally simple and remains close to a standard
Davidson procedure. For large systems, the main practical difference is memory
placement: when $\mathbf{T}_{ij}^P$, $\mathbf{T}_{ab}^P$, and the Davidson
subspace vectors $\mathbf{V}_{ia}^k$ and $\mathbf{W}_{ia}^k$ do not fit entirely
in GPU memory, they are kept in CPU memory and uploaded to the GPU only on
demand. In practice, the complete $\mathbf{V}$ and $\mathbf{W}$ subspaces are
required mainly during residual evaluation, Gram--Schmidt orthogonalization, and
reconstruction of the full excitation vectors $\mathbf{X}$. Thus, besides the
matrix--vector multiplication, an extra cost is CPU--GPU communication, bounded
by PCIE or NVLink bandwidth.

To maximize code reuse with modules already implemented in GPU4PySCF, all
two-center and three-center ERIs are generated in double precision (float64) and
then cast to single precision (float32). All subsequent TDDFT-risp operations,
including the Davidson diagonalization, are performed entirely in single
precision. When the accumulated Davidson subspace vectors approach the available
memory limit, the solver restarts: the current approximate eigenvectors are
retained as the initial guess for a fresh Krylov expansion, and the
corresponding $AV$ products are recomputed to preserve consistency. This
restart-on-memory-exhaustion strategy is a common technique in iterative
eigensolvers for large problems
and introduces no additional approximation beyond the finite subspace
truncation inherent to any restarted Krylov method.

\subsection{Computational Details}

\textbf{Hardware.}
Unless otherwise noted, all calculations are performed on a single NVIDIA
A100-80GB GPU. The host CPU is an Intel Xeon Platinum 8362 (2.80~GHz) with
256~GB RAM. For the CPU-assisted Davidson workflow, the K-term tensors and
Davidson vectors that exceed GPU memory are kept in host memory and
streamed to the GPU on demand.

\textbf{Software.}
GPU TDDFT-risp / TDA-risp calculations are performed using a developmental
version of GPU4PySCF\cite{wu2025wcms,li2025jpca} with the ris module, where python packages
\texttt{cupy} and \texttt{cutensor} are used to perform the matrix
multiplication and tensor contraction on GPU. The CPU version of the risp code
are built on PySCF~2.12\cite{sun2018wcms,sun2020jcp}, where \texttt{numpy} with
\texttt{mkl} backend is used. The CPU version has the same code design with GPU
code, the only difference is that the on-the-fly J engine is not supported on
the CPU. Reference conventional TDDFT calculations are performed with
ORCA~6.1.0\cite{neese2025wcms} using the \texttt{RIJCOSX DEFGRID2} defaults,
parallelized via \texttt{nprocs 32}. CPU-risp timings use
\texttt{OMP\_NUM\_THREADS=4}. Natural transition orbital
(NTO)\cite{martin2003jcp} and hole--electron analyses of ORCA results are
carried out with Multiwfn~\cite{lu2024jcp,lu2012jcc,liu2020c}. The corresponding
analyses for risp are performed with the \texttt{ris} module in GPU4PySCF.

\textbf{DFT settings.}
For EXTEST42 and large system, the ground-state DFT is calculated using
GPU4PySCF. The SCF convergence threshold is \texttt{conv\_tol}~$= 10^{-10}$ for
systems with less than 1000 atoms and $10^{-8}$ for the larger systems. The
function integration grid is the GPU4PySCF default. When comparing to ORCA
results, both GPU and CPU TDDFT-risp start from ORCA SCF orbitals to guarantee a
fair comparison, where MOKIT package is applied to transfer the orbitals from
ORCA to PySCF\cite{zou2024mokit}. There is no solvation model used in the
ground-state DFT calculation. Since we focus on the excited-state calculations,
the SCF procedure is not included in the timing profiling. We note that for Fock
matrix dimensions exceeding ${\sim}30\,000$, the GPU-accelerated
\texttt{cuSOLVER} eigensolver is unavailable due to its 32-bit integer API, and
the SCF diagonalization falls back to a CPU eigensolver. This becomes the
bottleneck cost for the large-system calculations. The full log files are
provided in the Zenodo\cite{zhou_2026_19237732}.

\textbf{Davidson settings.}
The residual norm convergence threshold varies by system size and convergence
behavior (see Table~\ref{tab:large_systems_profiling} and Supplementary
Information). Restarts are triggered when the accumulated expansion vectors
approach the available memory limit. The expansion vectors are naturally kept in
single precision, which introduces negligible numerical error. Full TDDFT-risp
involving (symplectic eigenvalue problem) is also solved by Davidson-like
algorithm, where both $(A+B)$ and $(A-B)$ are projected into the subspace
spanned by the corresponding expansion basis of $X+Y$ and $X-Y$.

\textbf{Benchmark protocol.}
The EXTEST42 accuracy benchmarks use
PBE0\cite{adamo1999jcp},
$\omega$B97XD\cite{chai2008pccp},
$\omega$B97X-D3BJ\cite{najibi2018jctc}, and
$\omega$B97MV\cite{mardirossian2016jcp} with the def2-TZVP
basis sets\cite{weigend2005pccp} (20 excited states).
Large-system profiling calculations
(Table~\ref{tab:large_systems_profiling}) employ
$\omega$B97XD/def2-SVP with 15 excited states and the 16~eV/no-H
setting. Wall-time comparisons with ORCA
(Table~\ref{tab:walltime_comparison}) use
$\omega$B97X-D3BJ/def2-TZVP with 20 excited states and the 40~eV/no-H
setting.

The root mean square error (RMSE) metric for excitation energies is defined as
\begin{equation}
\label{eq:rmse}
  \text{RMSE} = \left( \frac{ \sum_{i=1}^N  (\Omega_i^\text{ref}-\Omega_{j(i)})^2 } {N} \right)^{1/2} ,
\end{equation}
where $\Omega_i^\text{ref}$ is the $i$-th excitation energy computed using TDDFT,
$\Omega_{j(i)}$ is the corresponding excitation energy of the approximate method,
$N=20$ is the number of excited states considered and the
function $j(i)$ is related to the overlap of the TDDFT eigenvectors
and associates the excited state $i$ of the standard TDDFT calculation with the
excited state $j$ in the approximate calculation, thus avoiding state flipping
in the comparison of excitation energies. The association of states is defined as\cite{giannone2020jcp}
\begin{equation}
    O_{mn} = \left( \sum_{i,a} X_{ia}^{m,\mathrm{ref}} X_{ia}^{n} \right) \exp \left( - \left| f_m^{\mathrm{ref}} - f_n \right| \right)
\end{equation}

\section{Results and discussion}\label{sec:results}
\subsection{EXTEST42 Benchmark Accuracy}

In this subsection, we evaluate the excitation-energy accuracy on the EXTEST42
benchmark set with hybrid and RSH functionals. The key
adjustable parameter is the exchange-window threshold that controls the
trade-off between efficiency and accuracy.

\begin{figure}[!htbp]
    \centering
    \includegraphics[width=\linewidth]{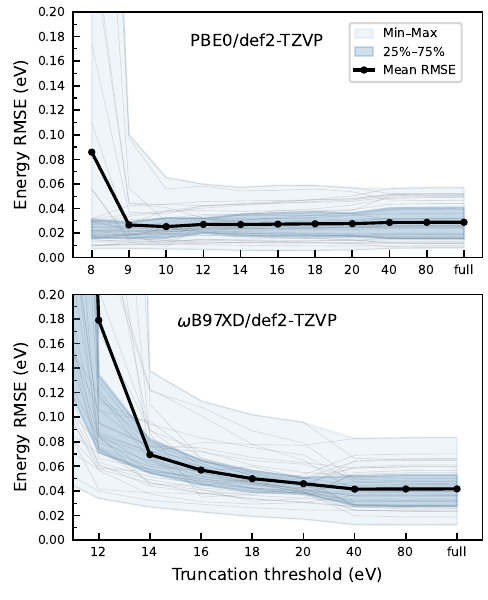}
    \caption{\label{fig:Ktrunc}
    Average 20-state excitation-energy RMSE (TDA-risp vs.\ standard TDA) on the
    EXTEST42 set as a function of the exchange-window truncation threshold.}
\end{figure}

Figure~\ref{fig:Ktrunc} shows the 20-state RMSE as a
function of the threshold for PBE0 and $\omega$B97XD with the def2-TZVP basis.
For PBE0, the RMSE remains around 0.03~eV even when the threshold is lowered
from 40~eV all the way to 10~eV, indicating that PBE0 is insensitive to
aggressive truncation. For $\omega$B97XD, the RMSE is 0.04~eV at 40~eV and rises
to 0.06~eV at 16~eV. The optimal threshold is functional-dependent, and additional
scans for other functionals and the def2-SVP basis are provided in the
Supporting Information.

Using the conservative 40~eV threshold, as listed in Figure~\ref{fig:rmse_mae}, the
PBE0 \ce{S_1} mean absolute error (MAE) is 0.029~eV with an average RMSE of
0.025~eV. For $\omega$B97XD, the corresponding values are 0.035~eV and 0.040~eV.
The full molecule-by-molecule data are given in the Supporting Information. For
the thousand-atom timings in Section~\ref{sec:walltime}, we adopt the more
aggressive 16~eV threshold to reduce the exchange cost, accepting a modestly
larger error. The spectral impact of this trade-off is examined in
Section~\ref{sec:spectral_validation}.

\subsection{Accuracy Impact of Hydrogen Exclusion}

\begin{figure}[!htbp]
    \centering
    \includegraphics[width=0.45\textwidth]{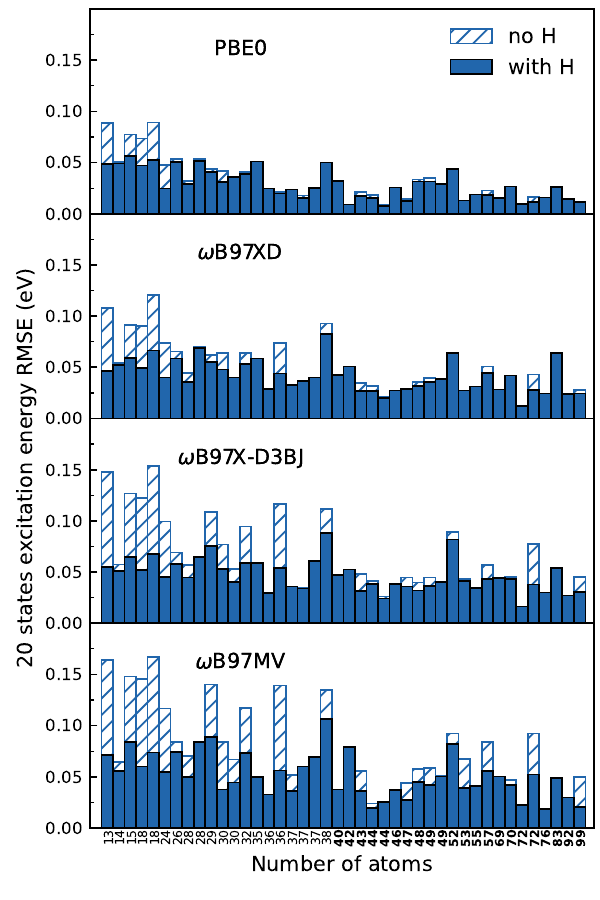}
    \caption{RMSE of the 20 lowest excitation energies (TDA-risp vs.\
    standard TDA) versus molecular size for the EXTEST42 set.}
    \label{fig:RMSE_trend}
\end{figure}

We next evaluate the accuracy impact of excluding hydrogen from the auxiliary
basis by computing the RMSE of the first 20 excited states on the
EXTEST42 set. Omitting hydrogen raises the RMSE appreciably for molecules with
$\leq$40 atoms but has little effect for $\geq$40 atoms. For
$\omega$B97X-D3BJ/def2-TZVP, molecules with $\leq$40 atoms typically have RMSE
above 0.05 eV, while those with $\geq$40 atoms mostly lie below 0.05 eV.
Figure~\ref{fig:RMSE_trend} shows, for each EXTEST42 molecule on def2-TZVP, the
RMSE of the 20 lowest excitation energies for TDA-risp relative to standard TDA,
for auxiliary bases constructed with and without hydrogen atoms.

\begin{figure}[!htbp]
    \centering
    \includegraphics[width=0.45\textwidth]{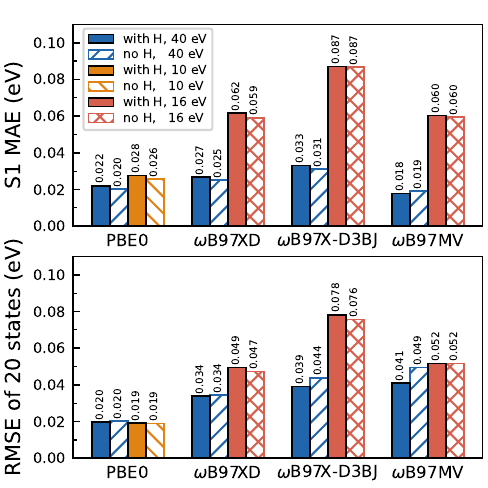}
    \caption{\label{fig:rmse_mae}
    Average \ce{S_1} Energy MAE and average RMSE of the 20 lowest excitations for
    TDA-risp on the EXTEST42 40--99~atom subset.}
\end{figure}

Figure~\ref{fig:rmse_mae} summarizes the TDA-risp excitation energy errors on the
40--99~atom molecules of EXTEST42, reported as the average mean absolute error
(MAE) of the \ce{S_1} excitation energy and the average RMSE
of the lowest 20 excited states, with def2-TZVP basis set. Values are averaged over
molecules for auxiliary bases with and without hydrogen atoms and for the
exchange truncation thresholds indicated in the figure.
Under the default 40~eV threshold with hydrogen included in the auxiliary basis,
all four functionals (PBE0, $\omega$B97XD, $\omega$B97X-D3BJ, and $\omega$B97MV)
yield an \ce{S_1} MAE below 0.04~eV and a
20-state RMSE below 0.05~eV.
Across every functional and truncation threshold tested,
removing hydrogen atoms from the auxiliary basis introduces at most
0.008~eV of additional error, confirming that hydrogen exclusion has a
negligible impact for molecules of this size.

Lowering the truncation threshold has a more functional-dependent effect. For
PBE0, the 10~eV threshold leaves the RMSE below 0.02~eV with only a modest
increase in \ce{S_1} MAE (from 0.022 to 0.028~eV). $\omega$B97XD and $\omega$B97MV
behave similarly at 16~eV: the \ce{S_1} MAE rises to about 0.06~eV and the RMSE
remains below 0.06~eV, indicating that a 16~eV threshold is a reasonable choice
for these functionals. $\omega$B97X-D3BJ is the most sensitive: at 16~eV the
\ce{S_1} MAE reaches 0.087~eV and the RMSE approaches 0.08~eV, suggesting that
16~eV is a rather aggressive truncation for this functional.

We emphasize that the lower thresholds (10 and 16~eV) are intended exclusively
for large systems where the exchange build dominates the wall time. For small
and medium-sized molecules the default 40~eV is already both fast and accurate.
The purpose of testing these aggressive thresholds here is to establish an upper
bound on the error, so that practitioners can make an informed choice when
pushing the method to its limits on very large systems.

\subsection{Spectral Validation}
\label{sec:spectral_validation}

\begin{figure*}[t]
    \centering
    \includegraphics[width=1\textwidth]{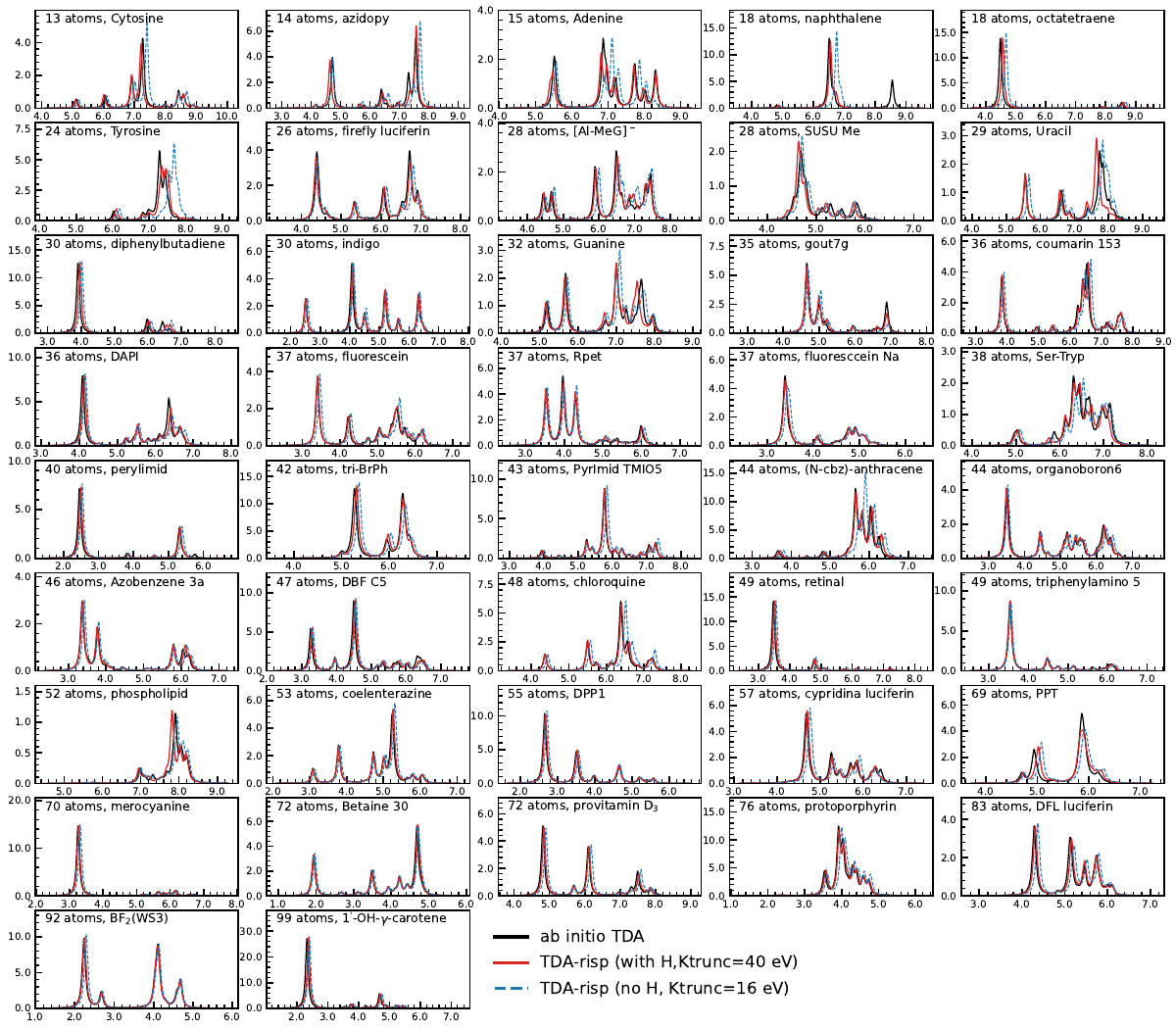}
    \caption{\label{fig:extest42_UV}
    UV--vis absorption spectra for all 42 molecules in the EXTEST42 set (20
    states, TDA-$\omega$B97XD/def2-TZVP, convergence tolerance = $1e-5$). Black:
    standard TDA reference. Red: TDA-risp with H in the auxiliary basis
    and a 40~eV truncation threshold (default setting). Blue dashed: TDA-risp
    without H in the auxiliary basis and a 16~eV truncation threshold
    (accelerated setting for large systems). Lorentzian broadening with HWHM = 0.05~eV.}
\end{figure*}

Having established the EXTEST42 accuracy of the exchange-window truncation and
hydrogen exclusion from the auxiliary basis, we next examine whether these scalar
error metrics translate into reliable spectral reproduction.
Figure~\ref{fig:extest42_UV} compares the UV--vis absorption spectra of all 42
EXTEST42 molecules under the conservative setting (40~eV, with H) and the most
aggressive production setting (16~eV, without H) against the standard TDA
reference. For molecules with $\geq$40 atoms, both settings produce spectra that
closely track the reference, indicating that the combined
approximation stack preserves not only average excitation energies but also
overall spectral shapes.

The aggressive 16~eV/no-H setting is designed specifically for large systems
where the full exchange build would otherwise be prohibitively expensive. The
EXTEST42 molecules themselves are small enough that the conservative
40~eV/with-H setting is already fast and accurate. Importantly, the spectral
deviations introduced by the aggressive setting diminish systematically with
increasing molecular size, consistent with the RMSE trend in
Figure~\ref{fig:RMSE_trend}. This size-dependent convergence gives confidence
that the aggressive setting becomes \emph{more} reliable, not less, as it is
applied to the thousand-atom systems for which it was intended. For small and
medium-sized molecules where computational cost is not a bottleneck, we
recommend the conservative 40~eV/with-H setting as the default choice.

\begin{figure*}[t]
    \centering
    \includegraphics[width=1\textwidth]{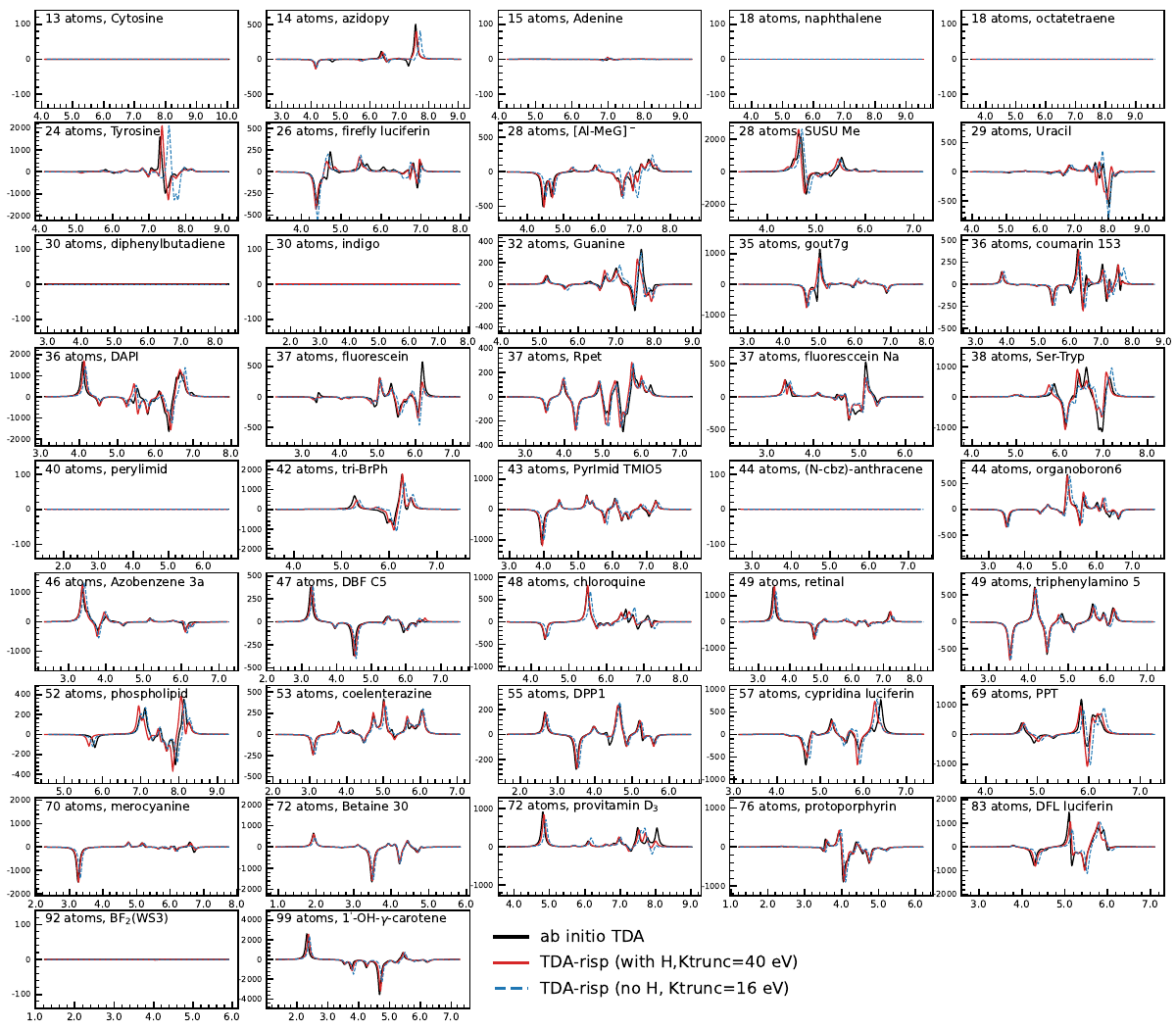}
    \caption{\label{fig:extest42_ECD}
    ECD spectra for all 42 molecules in the EXTEST42 set (20
    states, TDA-$\omega$B97XD/def2-TZVP, convergence tolerance = $1e-5$). Black:
    standard TDA reference. Red: TDA-risp with H in the auxiliary basis
    and a 40~eV truncation threshold (default setting). Blue dashed: TDA-risp
    without H in the auxiliary basis and a 16~eV truncation threshold
    (accelerated setting for large systems).}
\end{figure*}

Figure~\ref{fig:extest42_ECD} presents the corresponding ECD spectra for the
same 42 molecules and settings. Because ECD depends on the rotatory strength
Eq.~\eqref{eq:rotator_strength}, which involves the magnetic dipole transition
moment in addition to the electric dipole, it provides a more stringent test of
the approximation than UV--vis absorption alone. Under the default 40~eV/with-H
setting the ECD spectral shapes are well reproduced across the full set. The
accelerated 16~eV/no-H setting again shows visible deviations mainly for the
smallest molecules, while for molecules with $\geq$40 atoms the agreement
remains close.

\begin{figure}[!htbp]
    \centering
    \includegraphics[width=\linewidth]{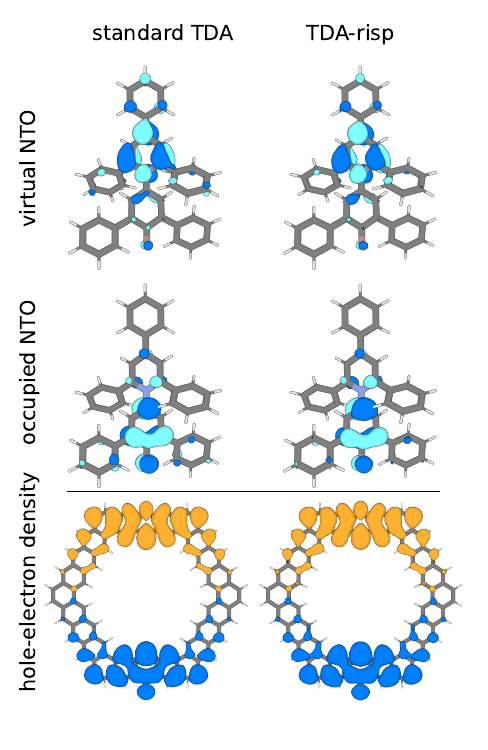}
    \caption{\label{fig:betaine30}
    NTO pair (singular value $>$ 0.9) of the \ce{S_1} state of Betaine~30
    molecule, blue and light blue stands for different orbital phases, isosurface 0.04. The
    hole-electron density analysis of the \ce{S_1} state of
    Donor-$\pi$-Acceptor model system, blue and yellow stands for hole and
    electron density, isosurface 0.0001. Calculated with $\omega$B97X-D3BJ/def2-TZVP.
    Standard TDA is calculated with ORCA 6.1.0.}
\end{figure}

As a qualitative check on charge-transfer (CT) states, we examine two
representative systems: Betaine~30 and a Donor-$\pi$-Acceptor model
molecule\cite{bane2025_daxitixi}. For Betaine~30, standard TDA followed by
NTO analysis reveals that the \ce{S_1} state is
dominated by a single NTO pair (singular value $>$ 0.9) with pronounced CT
character: the hole is localized on the $\ce{-O^-}$ moiety and its surrounding
atoms, while the electron is transferred to the \ce{N+}-containing region. For
the Donor-$\pi$-Acceptor system, hole--electron density analysis confirms the
same spatial separation, with the hole density concentrated on the donor
fragment and the electron density on the acceptor. As shown in
Figure~\ref{fig:betaine30}, TDA-risp faithfully reproduces both the NTO shapes
of Betaine~30 and the hole--electron density distribution of the
Donor-$\pi$-Acceptor molecule, even when hydrogen atoms are excluded from the
auxiliary basis. This agreement indicates that the risp approximation preserves
the long-range Coulomb and exchange coupling essential for describing CT
excitations, although a more systematic CT benchmark across diverse chromophore
classes would be desirable in future work.

\subsection{Large-System Timing Benchmarks}\label{sec:walltime}

We now assess the practical performance of TDA-risp on systems too large
for routine comparison with conventional reference calculations.
The primary objective is to demonstrate that the combined acceleration
strategy enables hybrid-functional excited-state calculations on a
single GPU at system sizes that would be prohibitive on CPU.

\subsubsection{Wall-Time Profiling}

The profiled large-system set spans several classes of systems, including
fluorescent proteins (PDBID 5EXB and 5EXC), polyacene aggregates
(cell1, cell2, cell3 and chain)\cite{einsele2025jpca}, OLED molecular
crystals (CCDC 2106623)\cite{liu2022jmcc}, and randomly packed aggregates of
the BODIPY dye 6a\cite{jimenez2021ic}, where 6a$N$ denotes an $N$-molecule aggregate
generated with Packmol\cite{martinez2009jcc}. All timings were obtained on a single NVIDIA
A100-80GB GPU and combine the three approximations validated above:
exchange-space truncation, hydrogen exclusion from the auxiliary basis, and the
CPU-assisted Davidson strategy. When the large tensors do not fit entirely in
GPU memory, the K-term tensors and Davidson $\mathbf{V}/\mathbf{W}$ vectors are
streamed from CPU memory in blocks, as described in the previous subsection.

Table~\ref{tab:large_systems_profiling} summarizes the wall-clock times for
these representative systems. It reports the memory usage (Mem) and wall time
for building $\mathbf{T}_{ij}^P$ and $\mathbf{T}_{ab}^P$, followed by the
Davidson memory footprint for $\mathbf{V}$ and $\mathbf{W}$, the wall time per
matrix--vector product (MVP), convergence tolerance, number of Davidson
iterations $N_\text{iter}$, number of restarts $N_\text{restart}$, and the
total wall time excluding the SCF step.
Representative systems in the 1000--3000 atom regime complete within practical
single-GPU wall times, ranging from tens of minutes for the easier cases to
several hours for the most demanding ones in Table~\ref{tab:large_systems_profiling}.
For example, the 3145-atom fluorescent protein 5EXC reaches convergence in about
100 min, whereas the more challenging 2770-atom COF example requires about
420 min. These results show that large-scale low-lying TDA-risp calculations
are feasible on a single A100 without multi-GPU or large CPU clusters.

\begin{table*}[t]
    \centering
    \caption{Profiling of TDA-risp $\omega$B97X-D/def2-SVP large-system
    calculations on a single NVIDIA A100-80GB GPU. All calculations use K-term
    truncation at 16 eV and request 15 excited states. J and K are the
    wall-clock times for Coulomb and exchange during one matrix-vector product.
    Total time records the wall-clock time for the linear reponse calculation,
    excluding the SCF step.}
    \label{tab:large_systems_profiling}
    {\small
    \begin{tabular}{>{\raggedright\arraybackslash}m{1.0cm}
                    >{\centering\arraybackslash}m{1.0cm}|
                    >{\centering\arraybackslash}m{1.0cm}|
                    >{\centering\arraybackslash}m{1.0cm}
                    >{\centering\arraybackslash}m{1.0cm}|
                    >{\centering\arraybackslash}m{1.0cm}
                    >{\centering\arraybackslash}m{0.8cm}   
                    >{\centering\arraybackslash}m{0.8cm}   
                    >{\centering\arraybackslash}m{1.0cm}
                    >{\centering\arraybackslash}m{0.7cm}
                    >{\centering\arraybackslash}m{1.0cm}|
                    >{\raggedleft\arraybackslash}m{1.0cm}@{\hspace{6mm}}}
    \toprule
    \multicolumn{3}{c}{} & \multicolumn{2}{c}{build $\mathbf{T}_{ij}^P$ $\mathbf{T}_{ab}^P$} & \multicolumn{6}{c}{Davidson diagonalization} \\
    \cmidrule(lr){4-5} \cmidrule(lr){6-11}
    System   & $N_\text{atoms}$ & ERI sparsity (\%) & Mem (GB)  & time (min)  & $\mathbf{V}$\&$\mathbf{W}$ (GB)  & J (sec) & K (sec) & conv tol & $N_\text{iter}$    & $N_\text{restart}$  & \multicolumn{1}{c}{\parbox[c]{1.8cm}{\centering total time \\ (min)}} \\
    \midrule
    chain    & 720      & 3    & 0.5       & 0.4       & 143       & 2        & 0        & 1e-04    & 43       & 0        & 6        \\
    cell1    & 660      & 9    & 0.5       & 0.3       & 70        & 5        & 0        & 1e-05    & 32       & 0        & 4        \\
    cell2    & 1320     & 5    & 3.6       & 2.1       & 201       & 20       & 4        & 1e-04    & 31       & 0        & 26       \\
    cell3    & 2070     & 3    & 28.7      & 10.4      & 170       & 64       & 89       & 1e-04    & 69       & 3        & 212      \\
    OLED1    & 1340     & 4    & 3.9       & 2.4       & 129       & 22       & 5        & 1e-04    & 60       & 1        & 50       \\
    OLED2    & 2680     & 2    & 31.3      & 23.3      & 163       & 134      & 88       & 5e-04    & 104      & 8        & 403      \\
    6a25     & 3000     & 1    & 44.9      & 27.1      & 149       & 131      & 143      & 1e-04    & 73       & 8        & 354      \\
    COF      & 2770     & 1    & 37.4      & 25.4      & 156       & 81       & 109      & 1e-03    & 110      & 15       & 442      \\
    5EXB     & 3145     & 2    & 47.4      & 25.9      & 129       & 124      & 158      & 1e-05    & 29       & 2        & 134      \\
    5EXC     & 3145     & 2    & 46.0      & 25.7      & 147       & 126      & 152      & 1e-05    & 19       & 1        & 110      \\
    \bottomrule
    \end{tabular}
    }
\end{table*}

Within the present validation scope, these timings show that
single-GPU TDA-risp calculations are feasible for low-lying excited states at
the thousand-atom scale.

\begin{figure}[!htbp]
    \centering
    \includegraphics[width=\linewidth]{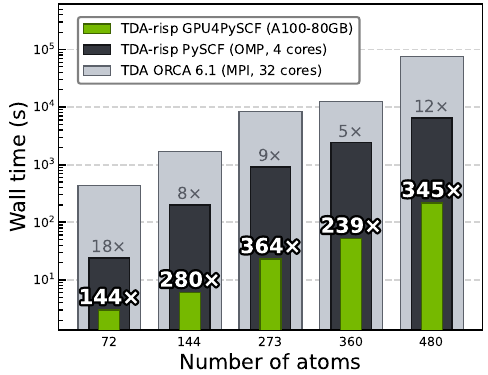}
    \caption{\label{fig:speedup}
    Wall-time speedup of TDA-risp on GPU (GPU4PySCF, single A100-80GB)
    and CPU (PySCF, 4 OpenMP threads) relative to RIJCOSX TDA in
    ORCA~6.1.0 (32 MPI processes). CPU Intel Xeon Platinum 8362.
    All calculations use $\omega$B97X-D3BJ/def2-TZVP with 20 excited
    states and the 40~eV/no-H risp setting.
   }
\end{figure}

For smaller systems where a conventional RIJCOSX reference is available,
Table~\ref{tab:walltime_comparison} compares the wall time of TDA-risp
(and TDDFT-risp) on a single A100 GPU (GPU4PySCF) and on CPU (PySCF,
4 OpenMP threads) against RIJCOSX TDA in ORCA~6.1.0 (32 MPI processes).
Figure~\ref{fig:speedup} summarizes the TDA speedups. On the smallest
system (Betaine~30, 72 atoms) GPU-risp is already $144\times$ faster than
ORCA and $8\times$ faster than CPU-risp. The advantage grows with system
size: for the 480-atom 6a4, GPU-risp achieves a $345\times$ speedup over
ORCA and a $30\times$ speedup over CPU-risp. Even CPU-risp running on
only 4 cores outperforms 32-core ORCA by $5$--$12\times$ across all
systems, confirming that the risp approximation itself accounts for the
bulk of the speedup, with GPU acceleration providing an additional order
of magnitude. The speedup of GPU-TDDFT-risp is shown in supplementary information.

\begin{figure}[!htbp]
    \centering
    \includegraphics[width=\linewidth]{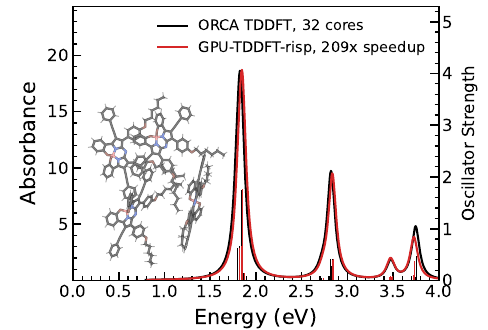}
    \caption{\label{fig:6a4_spectra}
    UV absorption spectrum of system 6a4 (480 atoms) computed with
    $\omega$B97X-D3BJ/def2-TZVP (20 states). The TDDFT-risp result
    (GPU4PySCF, single precision) is compared against conventional
    TDDFT (ORCA~6.1.0, RIJCOSX).}
\end{figure}

To verify that the combined risp approximations remain accurate beyond the
EXTEST42 benchmark set, we compare TDA-risp against conventional RIJCOSX TDDFT
(ORCA~6.1.0) on 6a4, a Packmol-generated tetramer of the BODIPY dye
6a\cite{jimenez2021ic} (480 atoms, $\omega$B97X-D3BJ/def2-TZVP, 20 states). The
\ce{S_1} excitation energy is 1.81~eV with TDDFT-risp versus 1.79~eV with ORCA,
an error of only 0.02~eV. As shown in Figure~\ref{fig:6a4_spectra}, the
broadened UV absorption spectra from the two methods are nearly
indistinguishable across the full 20-state energy range, confirming that the
risp approximation, including single-precision arithmetic, introduces negligible
spectral error even for a system well beyond the original benchmark size
range.

\begin{table*}[t]
    \centering
    \caption{Wall-time comparison of TDDFT-risp on GPU and CPU
    against RIJCOSX TDDFT (ORCA~6.1.0).
    All calculations use $\omega$B97X-D3BJ/def2-TZVP, 20 excited states,
    Ktrunc = 40~eV, no~H.}
    \label{tab:walltime_comparison}
    \begin{threeparttable}
    {\small
    \begin{tabular}{@{}l r| rrr| rrr@{}}
    \toprule
    \multicolumn{2}{c}{} & \multicolumn{6}{c}{Wall time (seconds)} \\
    \cmidrule(lr){3-8}
    \multicolumn{2}{c}{} & \multicolumn{3}{c}{TDA} & \multicolumn{3}{c}{TDDFT}\\
    \cmidrule(rl){3-5} \cmidrule(rl){6-8}
    System & $N_\text{atoms}$ & GPU-risp & CPU-risp \tnote{a} & ORCA\tnote{b} & GPU-risp & CPU-risp \tnote{a} & ORCA\tnote{b} \\
    \midrule
    Betaine 30   & 72       & 3        & 24       & 433      & 4        & 33       & 579      \\
    D-$\pi$-A    & 144      & 6        & 200      & 1,677    & 9        & 297      & 2,501    \\
    fluodb       & 273      & 23       & 919      & 8,380    & 40       & 1,675    & 18,424   \\
    6a3          & 360      & 52       & 2,420    & 12,430   & 111      & 4,625    & 38,083   \\
    6a4\tnote{c} & 480      & 217      & 6,445    & 74,822   & 750\tnote{d}& 14,120& 157,013  \\
    \bottomrule
    \end{tabular}
    }
    \begin{tablenotes}\footnotesize
    \item[a] PySCF 2.12.0, OMP parallel via \texttt{OMP\_NUM\_THREADS=4}.
    \item[b] ORCA 6.1.0, \texttt{RIJCOSX DEFGRID2} (default setup), MPI parallel via \texttt{nprocs 32}.
    \item[c] GPU code use on-the-fly J-engine to avoid 34 GB $\mathbf{T}_{ia}^P$ memory usage.
    \item[d] Projection vectors stored in CPU memory.
    \end{tablenotes}
    \end{threeparttable}
\end{table*}

\subsubsection{Davidson Algorithm in Large Systems}

\begin{figure}[!htbp]
    \centering
    \includegraphics[width=\linewidth]{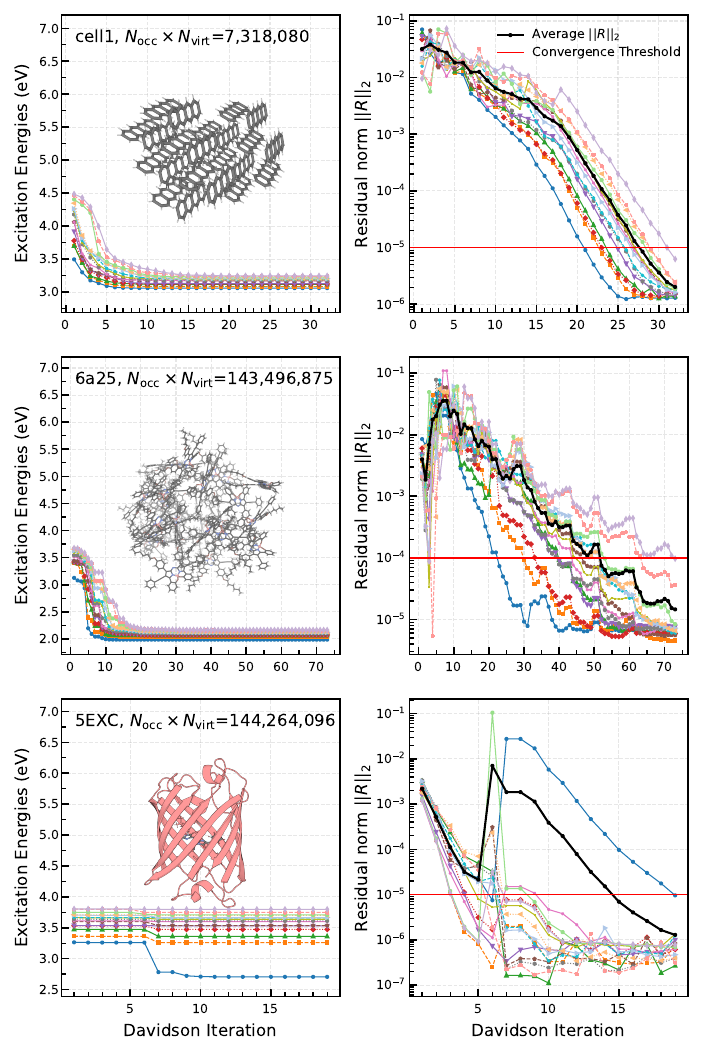}
    \caption{\label{fig:resplot}
    Davidson iteration history for 15 excited states
    (TDA-risp, $\omega$B97XD/def2-SVP) for cell1, 6a25, and
    5EXC. The black solid line is the average residual norm across
    all states.
   }
\end{figure}

A set of representative large-system examples further illustrates the
convergence behavior of the restarted Davidson procedure. As shown in
Figure~\ref{fig:resplot}, the left, middle, and right panels correspond to
cell1, 6a25, and 5EXC, respectively. For cell1, the average residual norm
decreases comparatively smoothly and reaches the $10^{-5}$ threshold after
27 iterations. The 6a25 aggregate is markedly more difficult: the
average residual oscillates throughout the first 10 iterations and even rises,
peaking at approximately $10^{-2}$ around iteration 8, before entering a more
steady decay regime and crossing the threshold only after about 45 iterations.
The 5EXC system shows a different but equally instructive pattern. Its average
residual drops rapidly during the first 6 iterations, giving the
impression of fast convergence, but after the first restart it increases
sharply and only then decreases again toward the $10^{-5}$ threshold near
iteration 20, while most individual states end up in the $10^{-6}$--$10^{-7}$
range.

Taken together, these examples show that Davidson convergence is substantially
more difficult for the larger systems than for the smaller ones. In particular,
an apparently favorable residual norm in the first few iterations does not
necessarily indicate robust convergence. The 5EXC case shows that continuing the
iterations can reveal latent instability that is not visible at the outset. We
therefore recommend carrying out at least 10 iterations for large systems before
judging the convergence behavior. More broadly, these observations highlight the
need for a more effective preconditioner and a better initial guess for
large-scale TDA-risp calculations. As reported in
Refs.~\cite{zhou2024jctc,helmich-paris2025jpca}, a well-engineered
preconditioner can speed up Davidson convergence by a factor of 2--3.

\subsection{Illustrative Large-System Excited-State Analysis}

Because direct spectral references are not readily available for the largest
systems considered here, we restrict the discussion below to qualitative
excited-state analyses intended to illustrate the type of information that can
still be extracted once such systems become computationally accessible.

Among the systems in Table~\ref{tab:large_systems_profiling}, the fluorescent
proteins provide a representative biomolecular example where a full quantum
treatment of the entire environment is desirable because the surrounding
$\beta$-barrel can significantly influence the chromophore
excitation~\cite{timerghazin2008jpcb}. We use the crystal structure of green
fluorescent protein (PDB ID: 5EXB)\cite{pletneva2016acd}. This structure can be
converted to the red form after $\sim$6 h photo-irradiation, and the
corresponding photoconverted structure 5EXC is also included in the test set.
Here we use fluorescent protein mainly as a representative biomolecular
large-system example for excited-state analysis, rather than for a detailed
mechanistic discussion.

For the excited-state analysis, we employ the hole--electron framework as
implemented in Multiwfn. This approach decomposes each electronic excitation
into a hole density (where the electron departs) and an electron density
(where it arrives), providing a direct real-space visualization of the
excitation character without relying on a single dominant NTO pair. When the
hole and electron densities overlap strongly, the excitation is local in nature.
When they are spatially separated, the excitation exhibits CT character.

\begin{figure}[!htbp]
    \centering
    \includegraphics[width=\linewidth]{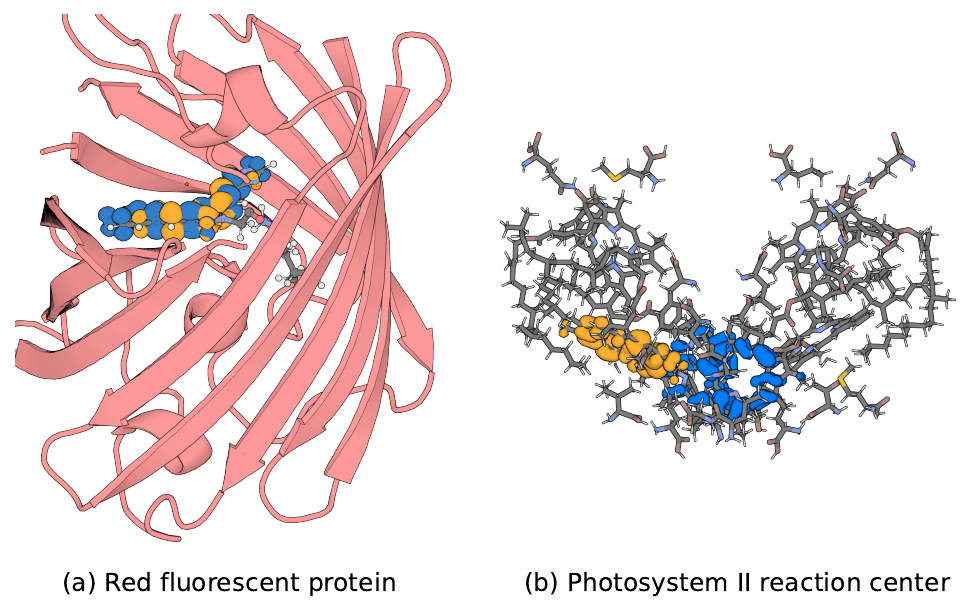}
    \caption{\label{fig:5exb_ps2}
    Hole--electron transition analysis of representative large-system excited
    states. (a) fluorescent protein 5EXC \ce{S_1},
    TDA-risp/$\omega$B97XD/def2-SVP, 3145 atoms. (b) PS~II reaction-center
    cluster S$_{20}$, TDDFT-risp/$\omega$B97XD/6-31G(dp), 1299 atoms. Blue and
    yellow isosurfaces denote the hole and electron densities, respectively.}
\end{figure}

Figure~\ref{fig:5exb_ps2} shows hole--electron analysis for two representative
large-system calculations. For fluorescent protein 5EXC, the \ce{S_1} state
displays a local excitation localized on the chromophore: both the hole and the
electron density are concentrated on the hydroxyphenyl--imidazolinone
$\pi$-system embedded in the $\beta$-barrel, with substantial spatial overlap,
consistent with a $\pi$--$\pi^*$ transition. Another example is Photosystem~II
(PS~II) reaction-center cluster (1299 atoms, TDDFT-risp/$\omega$B97XD/6-31G(dp),
Ktrunc=40 eV, no-H, SCF orbitals downloaded from Ref.~\cite{kavanagh2020pnas}),
the S$_{20}$ state exhibits Soret-band charge-transfer character: the hole
density is localized on one porphyrin ring while the electron density shifts
toward an adjacent chromophore, reflecting the inter-pigment coupling that
drives excitation-energy transfer in natural photosynthesis.

\section{Conclusion}
We have presented a GPU-accelerated implementation of TDDFT-risp (and TDA-risp)
in GPU4PySCF that combines a minimal $s/p$ auxiliary basis, on-the-fly AO-based
Coulomb evaluation, exchange-space truncation, hydrogen exclusion from the
auxiliary basis, and a host-memory-assisted Davidson solver. Together, these
make hybrid-functional excited-state calculations on a single A100 GPU practical
for systems of up to ${\sim}3000$ atoms.

On the EXTEST42 benchmark set, TDA-risp reproduces standard TDA excitation
energies with errors of 0.03--0.05~eV for low-lying states at a conservative
40~eV exchange cutoff, as confirmed by full UV--vis and ECD spectral
comparisons. With the more aggressive 16~eV threshold adopted for large
systems, the excited-state calculation alone takes tens of minutes to several
hours for 1000--3000 atoms and 15 excited states, achieving speedups of up to
$345\times$ over 32-core ORCA on a single GPU. Illustrative hole--electron
analyses of fluorescent protein and a photosystem~II model further demonstrate that the
method can provide physically meaningful excited-state characterization at
the thousand-atom scale.

The TDDFT-risp itself can, in principle, handle even larger systems than
reported here, but the ground-state SCF procedure becomes the practical
bottleneck. In particular, the eigenvalue routine in \texttt{cuSOLVER} (called
via CuPy) uses a 32-bit integer API, which limits the Fock matrix dimension to
${\sim}30\,000$. For Fock matrices exceeding this limit, the diagonalization
must fall back to a CPU eigensolver, which then dominates the overall wall time.

The present validation is limited to singlet excitations in closed-shell organic
molecules, and systems with proton-transfer-sensitive excited states may require
the more conservative default settings. Several extensions are natural next
steps: a more effective preconditioner and initial guess for the risp Davidson
solver; validation on open-shell species; generalization of the
hydrogen-exclusion strategy to systematically omit spectroscopically inactive
atoms from the auxiliary basis; application of these techniques to TDA-risp
analytical gradients for geometry optimization and dynamics; and parallelization
of the TDDFT-risp across multiple GPUs to go beyond ten thousands atoms.

\begin{acknowledgement}
    We thank Shane M. Parker for helpful discussions on TDDFT-risp and
    excited-state behavior in large systems. This work is supported by the Free
    Exploration Project (No.~XTS0018) of the Zhongguancun Academy.
\end{acknowledgement}

\begin{suppinfo}
    More details are available in the Supplementary Material. Code and data
    availability: TDDFT-risp modules are available in the open source code
    GPU4PySCF\cite{gpu4pyscf_github} and PySCF\cite{pyscf_github}. All benchmark
    data in this work are available on Zenodo\cite{zhou_2026_19237732}.
\end{suppinfo}

\bibliography{ris}

@ARTICLE{krause2017JCC,
  title = {Implementation of the bethe-salpeter equation in the TURBOMOLE program},
  author = {Krause, Katharina and Klopper, Wim},
  year = {2017},
  journal = {J. Comput. Chem.},
  volume = {38},
  number = {6},
  pages = {383--388},
  issn = {1096-987X},
  doi = {10.1002/jcc.24688},
  url = {https://onlinelibrary.wiley.com/doi/abs/10.1002/jcc.24688},
  urldate = {2022-11-22}
}

@article{kim2024jctc,
  title = {Very-{{Large-Scale GPU-Accelerated Nuclear Gradient}} of {{Time-Dependent Density Functional Theory}} with {{Tamm}}--{{Dancoff Approximation}} and {{Range-Separated Hybrid Functionals}}},
  author = {Kim, Inkoo and Jeong, Daun and Weisburn, Leah P. and Alexiu, Alexandra and Van Voorhis, Troy and Rhee, Young Min and Son, Won-Joon and Kim, Hyung-Jin and Yim, Jinkyu and Kim, Sungmin and Cho, Yeonchoo and Jang, Inkook and Lee, Seungmin and Kim, Dae Sin},
  year = 2024,
  month = oct,
  journal = {J. Chem. Theory Comput.},
  volume = {20},
  number = {20},
  pages = {9018--9031},
  publisher = {American Chemical Society},
  issn = {1549-9618},
  doi = {10.1021/acs.jctc.4c01003},
  urldate = {2025-12-01}
}

@article{craciunescu2023jctc,
  title = {Cluster-{{Based Approach Utilizing Optimally Tuned TD-DFT}} to {{Calculate Absorption Spectra}} of {{Organic Semiconductor Thin Films}}},
  author = {Craciunescu, Luca and Asbach, Maximilian and Wirsing, Sara and Hammer, Sebastian and Unger, Frederik and Broch, Katharina and Schreiber, Frank and Witte, Gregor and Dreuw, Andreas and Tegeder, Petra and Fantuzzi, Felipe and Engels, Bernd},
  year = 2023,
  month = dec,
  journal = {J. Chem. Theory Comput.},
  publisher = {American Chemical Society},
  doi = {10.1021/acs.jctc.3c01107},
  urldate = {2025-10-31},
  copyright = {\copyright{} 2023 American Chemical Society},
  langid = {english}
}

@article{timerghazin2008jpcb,
  title = {Computational {{Prediction}} of {{Absorbance Maxima}} for a {{Structurally Diverse Series}} of {{Engineered Green Fluorescent Protein Chromophores}}},
  author = {Timerghazin, Qadir K. and Carlson, Haley J. and Liang, Chen and Campbell, Robert E. and Brown, Alex},
  year = 2008,
  month = feb,
  journal = {J. Phys. Chem. B},
  volume = {112},
  number = {8},
  pages = {2533--2541},
  publisher = {American Chemical Society},
  issn = {1520-6106},
  doi = {10.1021/jp709900k},
  urldate = {2025-12-11}
}

@article{pletneva2016acd,
  title = {Crystal Structure of the Fluorescent Protein from {{Dendronephthya}} Sp. in Both Green and Photoconverted Red Forms},
  author = {Pletneva, N. V. and Pletnev, S. and Pakhomov, A. A. and Chertkova, R. V. and Martynov, V. I. and Muslinkina, L. and Dauter, Z. and Pletnev, V. Z.},
  year = 2016,
  month = aug,
  journal = {Acta Crystallogr., Sect. D: Struct. Biol.},
  volume = {72},
  number = {8},
  pages = {922--932},
  publisher = {International Union of Crystallography},
  issn = {2059-7983},
  doi = {10.1107/S205979831601038X},
  urldate = {2025-12-11},
  langid = {english}
}

@article{einsele2025jpca,
  title = {{{DIALECT}}, a {{Software Package}} for {{Exciton Spectra}} and {{Dynamics}} in {{Large Molecular Assemblies}} from {{Weak}} to {{Strong Light}}--{{Matter Coupling Regimes}}},
  author = {Einsele, Richard and Miao, Xincheng and Philipp, Luca Nils and Mitri{\'c}, Roland},
  year = 2025,
  month = jul,
  journal = {J. Phys. Chem. A},
  volume = {129},
  number = {30},
  pages = {6994--7007},
  publisher = {American Chemical Society},
  issn = {1089-5639},
  doi = {10.1021/acs.jpca.5c03307},
  urldate = {2025-11-11}
}

@article{athavale2025jctc,
  title = {{{PYSEQM}} 2.0: {{Accelerated Semiempirical Excited-State Calculations}} on {{Graphical Processing Units}}},
  author = {Athavale, Vishikh and Fedik, Nikita and Colglazier, William and Niklasson, Anders M. N. and Kulichenko, Maksim and Tretiak, Sergei},
  year = 2025,
  month = oct,
  journal = {J. Chem. Theory Comput.},
  volume = {21},
  number = {19},
  pages = {9498--9510},
  publisher = {American Chemical Society},
  issn = {1549-9618},
  doi = {10.1021/acs.jctc.5c00980}
}

@article{jimenez2021ic,
  title = {Tuning the {{Properties}} of {{Azadipyrromethene-Based Near-Infrared Dyes Using Intramolecular BO Chelation}} and {{Peripheral Substitutions}}},
  author = {Jimenez, Jayvic C. and Zhou, Zehao and Rheingold, Arnold L. and Parker, Shane M. and Sauv{\'e}, Genevi{\`e}ve},
  year = 2021,
  month = aug,
  journal = {Inorg. Chem.},
  pages = {13320--13331},
  publisher = {American Chemical Society},
  issn = {0020-1669},
  doi = {10.1021/acs.inorgchem.1c01597}
}

@article{ghosh2008JMS,
  title = {The Wave Mechanical Evaluation of the Absolute Radii of Atoms},
  author = {Ghosh, Dulal C. and Biswas, Raka and Chakraborty, Tanmoy and Islam, Nazmul and Rajak, Sandip K.},
  year = 2008,
  month = sep,
  journal = {J. Mol. Struct.: THEOCHEM},
  volume = {865},
  number = {1},
  pages = {60--67},
  issn = {0166-1280},
  doi = {10.1016/j.theochem.2008.06.020}
}

@article{sirohiwal2020jacs,
  title = {Protein {{Matrix Control}} of {{Reaction Center Excitation}} in {{Photosystem II}}},
  author = {Sirohiwal, Abhishek and Neese, Frank and Pantazis, Dimitrios A.},
  year = 2020,
  month = oct,
  journal = {J. Am. Chem. Soc.},
  volume = {142},
  number = {42},
  pages = {18174--18190},
  publisher = {American Chemical Society},
  issn = {0002-7863},
  doi = {10.1021/jacs.0c08526}
}

@article{bourneworster2024jcp,
  title = {Quantum Dynamics of Excited State Proton Transfer in Green Fluorescent Protein},
  author = {Bourne-Worster, Sam and Worth, Graham A.},
  year = 2024,
  journal = {J. Chem. Phys.},
  volume = {160},
  number = {6},
  pages = {065102},
  publisher = {AIP Publishing},
  doi = {10.1063/5.0188834}
}

@article{list2019jctc,
  title = {Assessment of Functionals for {{TDDFT}} Calculations of One- and Two-Photon Absorption Properties of Neutral and Anionic Fluorescent Protein Chromophores},
  author = {List, Nanna H. and Jensen, Frank and Kongsted, Jacob},
  year = 2019,
  journal = {J. Chem. Theory Comput.},
  volume = {15},
  number = {4},
  pages = {2492--2503},
  publisher = {American Chemical Society},
  doi = {10.1021/acs.jctc.8b00769}
}

@article{kavanagh2020pnas,
  title = {A {{TDDFT}} Investigation of the {{Photosystem II}} Reaction Center: {{Insights}} into the Precursors to Charge Separation},
  author = {Kavanagh, Maeve A. and Karlsson, Joshua K. G. and Colburn, Jonathan D. and Barter, Laura M. C. and Gould, Ian R.},
  year = 2020,
  month = aug,
  journal = {Proc. Natl. Acad. Sci. U.S.A.},
  volume = {117},
  number = {33},
  pages = {19705--19712},
  publisher = {Proceedings of the National Academy of Sciences},
  doi = {10.1073/pnas.1922158117}
}

@article{zhou2023jpcl,
  title = {Minimal Auxiliary Basis Set Approach for the Electronic Excitation Spectra of Organic Molecules},
  author = {Zhou, Zehao and Della Sala, Fabio and Parker, Shane M.},
  year = 2023,
  journal = {J. Phys. Chem. Lett.},
  volume = {14},
  number = {7},
  pages = {1968--1976},
  publisher = {American Chemical Society},
  doi = {10.1021/acs.jpclett.2c03698}
}

@article{zhou2024jctc,
  title = {Converging {{Time-Dependent Density Functional Theory Calculations}} in {{Five Iterations}} with {{Minimal Auxiliary Preconditioning}}},
  author = {Zhou, Zehao and Parker, Shane M.},
  year = 2024,
  journal = {J. Chem. Theory Comput.},
  volume = {20},
  number = {15},
  pages = {6738--6746},
  publisher = {American Chemical Society},
  issn = {1549-9618},
  doi = {10.1021/acs.jctc.4c00577}
}

@article{pu2026jctc,
  title = {Analytical {{Excited-State Gradients}} and {{Derivative Couplings}} in {{TDDFT}} with {{Minimal Auxiliary Basis Set Approximation}} and {{GPU Acceleration}}},
  author = {Pu, Zhichen and Wu, Xiaojie and Wang, Yuanheng and Fan, Cheng and Yan, Wen and Zhou, Zehao and Gao, Yi Qin and Sun, Qiming},
  year = 2026,
  journal = {J. Chem. Theory Comput.},
  publisher = {American Chemical Society},
  issn = {1549-9618},
  doi = {10.1021/acs.jctc.5c01960}
}

@article{lv2026c,
  title = {Generalized {{Energy-Based Fragmentation TDDFT-ris Method}} and Its {{Application}} to {{Electronic Spectra}} and {{Asymmetry Factors}} of {{Large Systems}}},
  author = {Lv, Ziyi and Wang, Fengmin and Sun, Junhui and Wang, Xuerong and Li, Shuhua and Zhou, Zehao and Zou, Jingxiang and Li, Wei},
  year = 2026,
  month = mar,
  journal = {Chin. J. Chem. Phys.},
  publisher = {2016 CPS},
  issn = {1674-0068}
}

@article{helmich-paris2025jpca,
  title = {A {{Two-Level Preconditioner}} for the {{CASSCF Linear-Response Equations}}},
  author = {Helmich-Paris, Benjamin},
  year = 2025,
  journal = {J. Phys. Chem. A},
  volume = {129},
  number = {35},
  pages = {8228--8238},
  publisher = {American Chemical Society},
  issn = {1089-5639},
  doi = {10.1021/acs.jpca.5c04385}
}

@article{giannone2020jcp,
  title = {Minimal Auxiliary Basis Set for Time-Dependent Density Functional Theory and Comparison with Tight-Binding Approximations: {{Application}} to Silver Nanoparticles},
  author = {Giannone, Giulia and Della Sala, Fabio},
  year = 2020,
  month = aug,
  journal = {J. Chem. Phys.},
  volume = {153},
  number = {8},
  pages = {084110},
  publisher = {American Institute of Physics},
  issn = {0021-9606},
  doi = {10.1063/5.0020545}
}

@article{liu2022jmcc,
  title = {Crystalline Organic Thin Films for Crystalline {{OLEDs}} ({{I}}): Orientation of Phenanthroimidazole Derivatives},
  author = {Liu, Dan and Zhu, Feng and Yan, Donghang},
  year = 2022,
  month = feb,
  journal = {J. Mater. Chem. C},
  volume = {10},
  number = {7},
  pages = {2663--2670},
  publisher = {The Royal Society of Chemistry},
  issn = {2050-7534},
  doi = {10.1039/D1TC04286F}
}

@misc{bane2025_daxitixi,
  author       = {Wei, Chiyuan},
  title        = {Schemes for Excited-State Wavefunction Analysis in Large Systems},
  url          = {https://bane-dysta.github.io/posts/47/},
  year         = {2025},
  month        = {aug},
}

@article{lu2012jcc,
  title = {Multiwfn: {{A}} Multifunctional Wavefunction Analyzer},
  author = {Lu, Tian and Chen, Feiwu},
  year = 2012,
  journal = {J. Comput. Chem.},
  volume = {33},
  number = {5},
  pages = {580--592},
  issn = {1096-987X},
  doi = {10.1002/jcc.22885}
}

@article{lu2024jcp,
  title = {A Comprehensive Electron Wavefunction Analysis Toolbox for Chemists, {{Multiwfn}}},
  author = {Lu, Tian},
  year = 2024,
  journal = {J. Chem. Phys.},
  volume = {161},
  number = {8},
  pages = {082503},
  issn = {0021-9606},
  doi = {10.1063/5.0216272}
}

@article{liu2020c,
  title = {An Sp-Hybridized All-Carboatomic Ring, Cyclo[18]Carbon: {{Electronic}} Structure, Electronic Spectrum, and Optical Nonlinearity},
  author = {Liu, Zeyu and Lu, Tian and Chen, Qinxue},
  year = 2020,
  journal = {Carbon},
  volume = {165},
  pages = {461--467},
  issn = {0008-6223},
  doi = {10.1016/j.carbon.2020.05.023}
}

@article{martinez2009jcc,
  title = {{{PACKMOL}}: {{A}} Package for Building Initial Configurations for Molecular Dynamics Simulations},
  author = {Mart{\'i}nez, L. and Andrade, R. and Birgin, E. G. and Mart{\'i}nez, J. M.},
  year = 2009,
  journal = {J. Comput. Chem.},
  volume = {30},
  number = {13},
  pages = {2157--2164},
  doi = {10.1002/jcc.21224}
}

@article{grimme2013simplified,
  title = {A Simplified Tamm--Dancoff Density Functional Approach for the Electronic Excitation Spectra of Very Large Molecules},
  author = {Grimme, Stefan},
  year = 2013,
  journal = {J. Chem. Phys.},
  volume = {138},
  number = {24},
  pages = {244104},
  publisher = {AIP Publishing},
  doi = {10.1063/1.4811331}
}

@article{bannwarth2014CaTCa,
  title = {A Simplified Time-Dependent Density Functional Theory Approach for Electronic Ultraviolet and Circular Dichroism Spectra of Very Large Molecules},
  author = {Bannwarth, Christoph and Grimme, Stefan},
  year = 2014,
  journal = {Comput. Theor. Chem.},
  series = {Excited States: {{From}} Isolated Molecules to Complex Environments},
  volume = {1040--1041},
  pages = {45--53},
  issn = {2210-271X},
  doi = {10.1016/j.comptc.2014.02.023}
}

@article{grimme2016jcp,
  title = {Ultra-Fast Computation of Electronic Spectra for Large Systems by Tight-Binding Based Simplified {{Tamm-Dancoff}} Approximation ({{sTDA-xTB}})},
  author = {Grimme, Stefan and Bannwarth, Christoph},
  year = 2016,
  month = aug,
  journal = {J. Chem. Phys.},
  volume = {145},
  number = {5},
  pages = {054103},
  issn = {0021-9606},
  doi = {10.1063/1.4959605}
}

@article{zhang2025jctcb,
  title = {The {{Atomic Density-Based Tight-Binding}} ({{aTB}}) {{Model}}: {{A Robust}} and {{Accurate Semiempirical Method Parametrized}} for {{H}}--{{Ra}}; {{Applied}} to {{Structures}}, {{Vibrational Frequencies}}, {{Noncovalent Interactions}}, and {{Excited States}}},
  author = {Zhang, Yingfeng and Xiao, Jin and Wang, Shunyu and Zhu, Tong and Zhang, John Z. H.},
  year = 2025,
  month = apr,
  journal = {J. Chem. Theory Comput.},
  volume = {21},
  number = {7},
  pages = {3410--3425},
  publisher = {American Chemical Society},
  issn = {1549-9618},
  doi = {10.1021/acs.jctc.4c01694}
}

@incollection{casida1995,
  title = {Time-Dependent Density Functional Response Theory for Molecules},
  author = {Casida, Mark E.},
  year = 1995,
  booktitle = {Recent Advances in Density Functional Methods},
  publisher = {World Scientific},
  address = {Singapore},
  pages = {155--192},
  doi = {10.1142/9789812830586_0005}
}

@article{hirata1999cpl,
  title = {Time-Dependent Density Functional Theory within the {{Tamm--Dancoff}} Approximation},
  author = {Hirata, So and Head-Gordon, Martin},
  year = 1999,
  journal = {Chem. Phys. Lett.},
  volume = {314},
  pages = {291--299},
  doi = {10.1016/S0009-2614(99)01149-5}
}

@article{martin2003jcp,
  title = {Natural Transition Orbitals},
  author = {Martin, Richard L.},
  year = 2003,
  journal = {J. Chem. Phys.},
  volume = {118},
  pages = {4775--4777},
  doi = {10.1063/1.1558471}
}

@article{adamo1999jcp,
  title = {Toward Reliable Density Functional Methods without Adjustable Parameters: {{The PBE0}} Model},
  author = {Adamo, Carlo and Barone, Vincenzo},
  year = 1999,
  journal = {J. Chem. Phys.},
  volume = {110},
  pages = {6158--6170},
  doi = {10.1063/1.478522}
}

@article{chai2008pccp,
  title = {Long-Range Corrected Hybrid Density Functionals with Damped Atom--Atom Dispersion Corrections},
  author = {Chai, Jeng-Da and Head-Gordon, Martin},
  year = 2008,
  journal = {Phys. Chem. Chem. Phys.},
  volume = {10},
  pages = {6615--6620},
  doi = {10.1039/B810189B}
}

@article{najibi2018jctc,
  title = {{{DFT-D3(BJ)}} Variants of the {{B97M-V}}, {$\omega$}{{B97X-V}} and {$\omega$}{{B97M-V}} Functionals},
  author = {Najibi, Asim and Goerigk, Lars},
  year = 2018,
  journal = {J. Chem. Theory Comput.},
  volume = {14},
  pages = {5725--5738},
  doi = {10.1021/acs.jctc.8b00842}
}

@article{mardirossian2016jcp,
  title = {{$\omega$}{{B97M-V}}: {{A}} Combinatorially Optimized, Range-Separated Hybrid, Meta-{{GGA}} Density Functional with {{VV10}} Nonlocal Correlation},
  author = {Mardirossian, Narbe and Head-Gordon, Martin},
  year = 2016,
  journal = {J. Chem. Phys.},
  volume = {144},
  pages = {214110},
  doi = {10.1063/1.4952647}
}

@article{weigend2005pccp,
  title = {Balanced Basis Sets of Split Valence, Triple Zeta Valence and Quadruple Zeta Valence Quality for {{H}} to {{Rn}}: {{Design}} and Assessment of Accuracy},
  author = {Weigend, Florian and Ahlrichs, Reinhart},
  year = 2005,
  journal = {Phys. Chem. Chem. Phys.},
  volume = {7},
  pages = {3297--3305},
  doi = {10.1039/B508541A}
}

@article{sun2018wcms,
  title = {{{PySCF}}: The {{Python}}-Based Simulations of Chemistry Framework},
  author = {Sun, Qiming and Berkelbach, Timothy C. and Blunt, Nick S. and Booth, George H. and Guo, Sheng and Li, Zhendong and Liu, Junzi and McClain, James D. and Sayfutyarova, Elvira R. and Sharma, Sandeep and Wouters, Sebastian and Chan, Garnet Kin-Lic},
  year = 2018,
  journal = {WIREs Comput. Mol. Sci.},
  volume = {8},
  pages = {e1340},
  doi = {10.1002/wcms.1340}
}

@article{sun2020jcp,
  title = {Recent Developments in the {{PySCF}} Program Package},
  author = {Sun, Qiming and Zhang, Xing and Banerjee, Samragni and Bao, Peng and Barbry, Marc and Blunt, Nick S. and Bogdanov, Nikolay A. and Booth, George H. and Chen, Jia and Cui, Zhi-Hao and others},
  year = 2020,
  journal = {J. Chem. Phys.},
  volume = {153},
  pages = {024109},
  doi = {10.1063/5.0006074}
}

@article{wu2025wcms,
  title = {Enhancing {{GPU}}-Acceleration in the {{Python}}-Based Simulations of Chemistry Frameworks},
  author = {Wu, Xiaojie and Sun, Qiming and Pu, Zhichen and Zheng, Tianze and Ma, Wenzhi and Yan, Wen and Xia, Yu and Wu, Zhengxiao and Huo, Mian and Li, Xiang and Ren, Weiluo and Gong, Sheng and Zhang, Yumin and Gao, Weihao},
  year = 2025,
  journal = {WIREs Comput. Mol. Sci.},
  volume = {15},
  pages = {e70008},
  doi = {10.1002/wcms.70008}
}

@article{li2025jpca,
  title = {Introducing {{GPU Acceleration}} into the {{Python-Based Simulations}} of {{Chemistry Framework}}},
  author = {Li, Rui and Sun, Qiming and Zhang, Xing and Chan, Garnet Kin-Lic},
  year = 2025,
  month = feb,
  journal = {The Journal of Physical Chemistry A},
  volume = {129},
  number = {5},
  pages = {1459--1468},
  publisher = {American Chemical Society},
  issn = {1089-5639},
  doi = {10.1021/acs.jpca.4c05876}
}

@misc{gpu4pyscf_github,
  title = {{GPU4PySCF}},
  url = {https://github.com/pyscf/gpu4pyscf},
  note = {GitHub repository}
}

@misc{pyscf_github,
  title = {{PySCF}},
  url = {https://github.com/pyscf/pyscf},
  note = {GitHub repository}
}

@article{neese2025wcms,
  title = {Software Update: {{The ORCA}} Program System---Version 6.0},
  author = {Neese, Frank},
  year = 2025,
  journal = {WIREs Comput. Mol. Sci.},
  volume = {15},
  pages = {e70019},
  doi = {10.1002/wcms.70019}
}

@article{ruger2016jcp,
  title = {Tight-Binding Approximations to Time-Dependent Density Functional Theory --- {{A}} Fast Approach for the Calculation of Electronically Excited States},
  author = {R{\"u}ger, Robert and {van Lenthe}, Erik and Heine, Thomas and Visscher, Lucas},
  year = 2016,
  month = may,
  journal = {J. Chem. Phys.},
  volume = {144},
  number = {18},
  pages = {184103},
  issn = {0021-9606},
  doi = {10.1063/1.4948647}
}

@misc{zou2024mokit,
  author = {Zou, Jingxiang},
  title = {{MOKIT} Program},
  url = {https://gitlab.com/jxzou/mokit},
  year = 2024,
  note = {accessed Apr 13, 2024}
}

@misc{zhou_2026_19237732,
  author       = {Zhou, Zehao},
  title        = {Raw Data for "GPU Accelerated Minimal Auxiliary
                   Basis Approach TDDFT for Large Organic Molecules"
                  },
  month        = mar,
  year         = 2026,
  publisher    = {zenedo},
  doi          = {10.5281/zenodo.19237732},
  url          = {https://doi.org/10.5281/zenodo.19237732},
}

\end{document}